\documentclass{article}

\usepackage{arxiv}

\usepackage[utf8]{inputenc} 
\usepackage[T1]{fontenc}    
\usepackage{hyperref}       
\usepackage{url}            
\usepackage{booktabs}       
\usepackage{amsfonts}       
\usepackage{nicefrac}       
\usepackage{microtype}      
\usepackage{lipsum}
\usepackage{graphicx}
\usepackage{amsmath}
\usepackage{subcaption} 
\usepackage{array}
\usepackage{tabularx}
\usepackage{placeins}
\usepackage[nottoc]{tocbibind}
\usepackage{hyperref}
\usepackage[table]{xcolor}
\usepackage{graphicx}
\usepackage{booktabs}
\usepackage{multirow}
\usepackage{lscape}
\usepackage{geometry}
\usepackage{authblk}
\usepackage{siunitx}
\definecolor{lightgray}{gray}{0.9}
\renewcommand{\arraystretch}{1.2}
\usepackage{longtable}
\graphicspath{ {./images/} }
\hypersetup{
  colorlinks=true,
  linkcolor=blue,     
  citecolor=green     
}
\title{Devising PoPStat: A metric bridging population pyramids with global disease mortality}
\author[1]{Tharaka Fonseka (BSc Eng.)}
\author[2]{Buddhi Wijenayake (BSc Eng.)}
\author[2]{Athulya Ratnayake (BSc Eng.)}
\author[3]{Inosha Alwis (MSc.)}
\author[4]{Supun Manathunga (MBBS)}
\author[2]{Roshan Godaliyadda (PhD)}
\author[5]{Samath Dharmarathne (MD)}
\author[2]{Vijitha Herath (PhD)}
\author[2]{Parakrama Ekanayake (PhD)}
\author[2]{Isuru Pamuditha (BSc Eng.)}

\affil[1]{Multidisciplinary AI Research Center, University of Peradeniya, Peradeniya, Sri Lanka}
\affil[2]{Department of Electrical and Electronic Engineering, Faculty of Engineering, University of Peradeniya, Peradeniya, Sri Lanka}
\affil[3]{Australian Centre for Health Services Innovation (AusHSI) and Centre for Healthcare
Transformation, School of Public Health and Social Work, Queensland University of
Technology (QUT), Brisbane, Queensland, Australia}
\affil[4]{Division of Experimental Medicine, Department of Medicine, Faculty of Medicine, McGill University, Canada}
\affil[5]{Department of Community Medicine, Faculty of Medicine, University of Peradeniya, Peradeniya, Sri Lanka}

\begin{document}
\maketitle
\vspace{2cm}

\vspace{1cm}
{\large\textbf{{Abstract}}}

{\textcolor{purple}{Background}}
Traditional demographic indicators offer a limited view of a country’s population structure and may not capture demographic influences on mortality patterns comprehensively. We aimed to bridge this gap by developing novel scalar metrics that condense the information in population pyramids and assess their association with disease specific mortality.

{\textcolor{purple}{Methods}}
Country specific population pyramids were constructed using the United Nations World Population Prospects 2024, while mortality data for 371 diseases across 180 countries were extracted from the Global Burden of Disease Study 2021. We then developed two metrics: PoPDivergence, which quantifies the difference between a country’s population pyramid and an optimized reference pyramid using Kullback Leibler divergence, and PoPStat, which is the correlation between PoPDivergence and cause specific mortality rates.

{\textcolor{purple}{Findings}}
 Non communicable diseases (NCDs) showed a strong PoPStat of –0.84 (optimized reference: Japan, p<0.001) with mortality concentrated to constrictive pyramids. Communicable, maternal, neonatal, and nutritional diseases had a moderate PoPStat of 0.50 (Singapore, p<0.001) with mortality linked to expansive pyramids. Injuries exhibited a weak PoPStat (0.291, Singapore, < 0.001). In more granular analyses, NCDs like neurological disorders and neoplasms, communicable diseases (CDs) like neglected tropical diseases, and other infections showed a strong PoPStat. Nevertheless, NCDs like diabetes, cirrhosis, and other chronic liver diseases, CDs like respiratory infections and tuberculosis carried a weak PoPStat indicating minimal influence of population pyramid on their mortality.

{\textcolor{purple}{Interpretations}}
This study, its devised metrics, demonstrates the degree to which the mortality of different diseases is bound to the underlying population structure and reveals what type of population pyramids will carry the highest mortality attributed to those diseases. 

{\textcolor{purple}{Funding}}
This study received no external funding. 

{\textcolor{purple}{Keywords}}
Population Pyramids, Demographic Transition, Epidemiology, PopDivergence, PoPStat

{\textcolor{purple}{Codespace}}
The full source code and analytic scripts for this study are available in the GitHub Codespace repository at \url{https://github.com/Buddhi19/DevisingPoPStat.git}.

\fbox{\parbox{0.98\textwidth}{
    \subsection*{\textcolor{purple}{Research in context}}
    \subsubsection*{Evidence before this study}
      To assess the current state of evidence, we systematically searched PubMed, Scopus, and Google Scholar on December 5, 2024, using different combinations of the keywords, “mortality”, “population pyramid”, “median age”, “life expectancy”, "demographic" and “correlation” with no restrictions placed on language or publication date. Previous research has studied associations between well known demographic indicators, such as median age, population density, and life expectancy, and various causes of mortality. In relation to population pyramids, we identified only narrative syntheses that assessed their impact on cause specific mortality. Most quantitative work on population pyramids have focused on clustering patterns and population projections. There remains a significant research gap in developing scalar metrics that can robustly quantify the influence of the entire population pyramid shapes on mortality outcomes.
    \subsubsection*{Added value of this study}
        This study was a novel attempt to quantitatively analyze the association between the shape of population pyramids and the type of mortality causes. We do so by constructing population pyramids from the most recently available population datasets and utilizing mortality data from a database with comprehensive coverage.  We introduced two new metrics, PoPDivergence and PoPStat, that broaden our understanding of demographic determinants of disease specific mortality. They also offer empirical support to the demographic and epidemiological transition models.
    \subsubsection*{Implications of all the available evidence}
    By summarizing information rich population pyramids into a scalar metric PoPDivergence opens a new avenue in research to assess association of any social, behavioural or economic variable with population pyramid morphology. The ability of PoPDivergence to locate stage of each country in the demographic transition allows it to capture population change, that happens across generations, within a single time point. This facilitates research on demographic transition without requiring longitudinal data over decades. Meanwhile, PoPStat provides a framework in epidemiology to study how underlying demographic structure impacts mortality of different diseases. As implications for practice, at the national level, PoPStat enables reliable predictions on disease specific mortality using population pyramid shapes, in contexts where mortality data may be missing or insufficient for a specific disease. Our findings also inform health policymakers of the types of diseases they need to prioritize given the stage of demographic transition they are in, which can guide resource allocation and planning in health. For global health, these metrics can introduce a novel at risk approach among countries based on the vulnerabilities of their population pyramid shapes for specific diseases.
}}
\section{Introduction}

Globally, approximately 60 million deaths occur per year, with their causes often being tied to socioeconomic disparities\cite{noauthor_deaths_nodate}. Understanding social and economic determinants of mortality is therefore crucial for developing effective healthcare policies and preventive interventions. Consequently, many studies have explored factors influencing mortality variations, with socioeconomic indicators like income level, Socio Demographic Index (SDI), and Human Development Index (HDI) being central to understanding mortality patterns\cite{G_Wang_2017,G_Bray_2012}. Additional social determinants, such as access to healthcare, employment conditions and living standards, have also been studied extensively for their impact on mortality outcomes\cite{M_Fullman_2018,W_Frank_2023}.

\begin{figure}[htp]
  \centering
  \begin{subfigure}{0.3\textwidth}
    \centering
    \includegraphics[width=\linewidth]{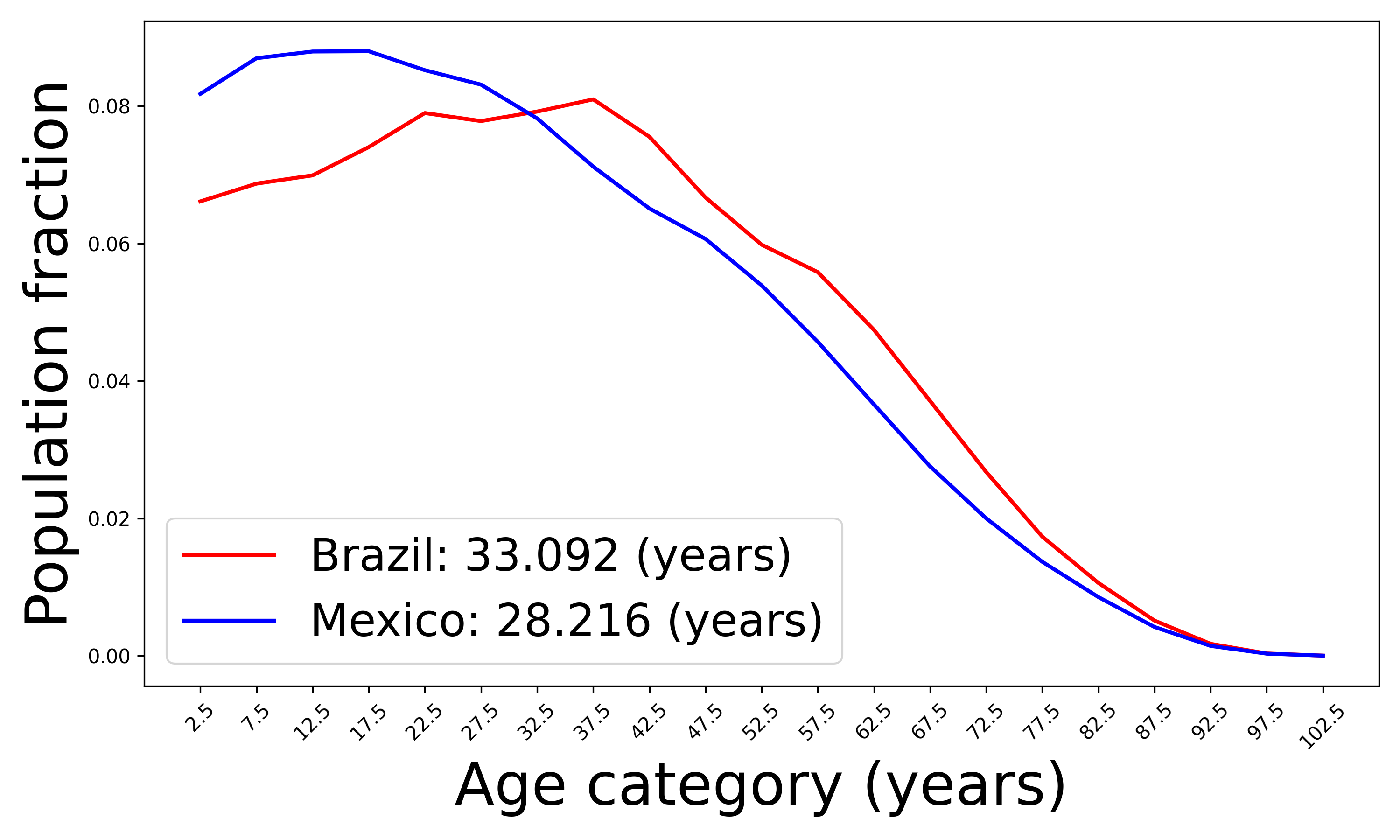}
    \caption{Difference in median age for similar population distribution.}
    \label{fig3:median_structure}
  \end{subfigure}%
  \hspace{1cm}
  \begin{subfigure}{0.3\textwidth}
    \centering
    \includegraphics[width=\linewidth]{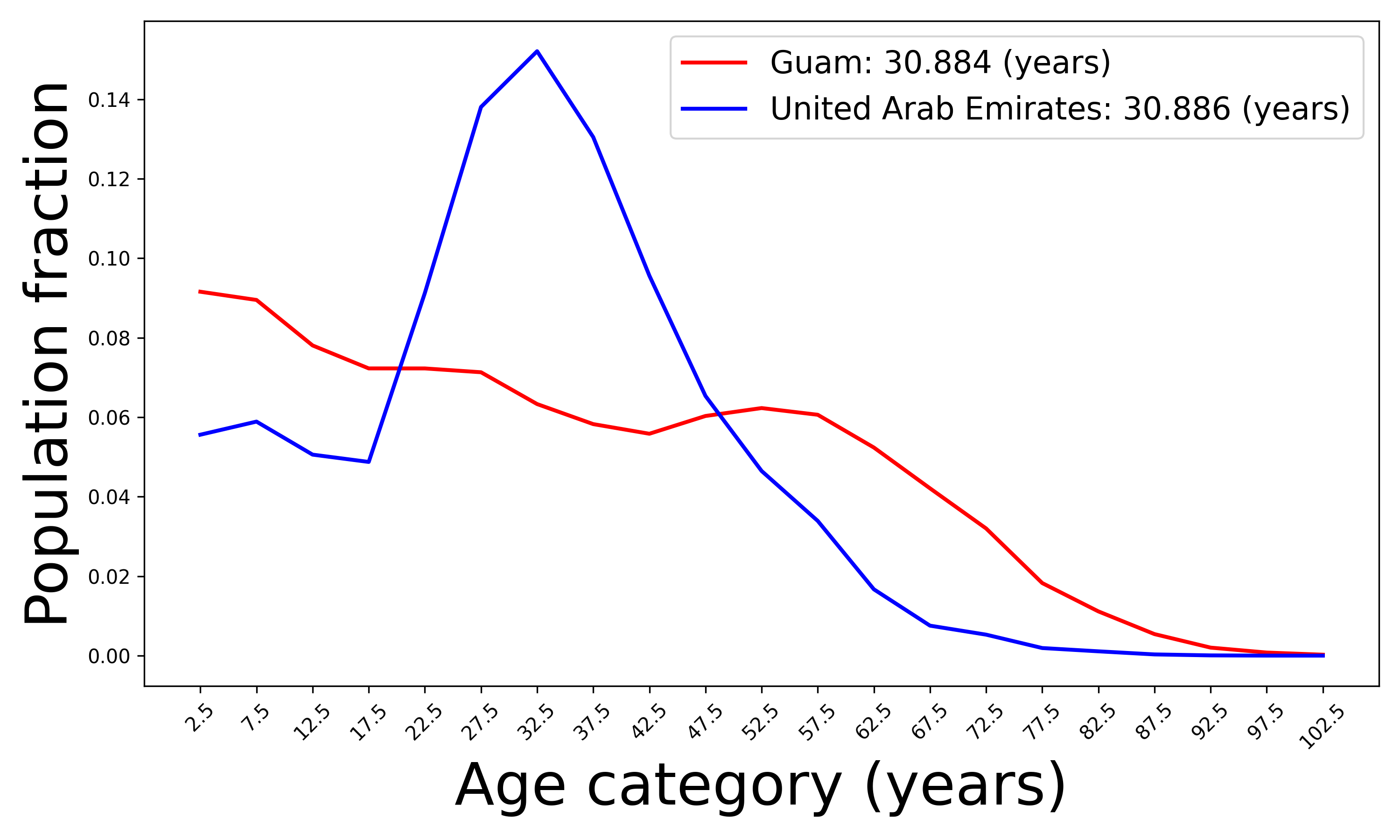}
    \caption{Difference in population distribution for similar median age.}
    \label{fig3:median_index}
  \end{subfigure}

    \vspace{0.5cm}

  \begin{subfigure}{0.3\textwidth}
    \centering
    \includegraphics[width=\linewidth]{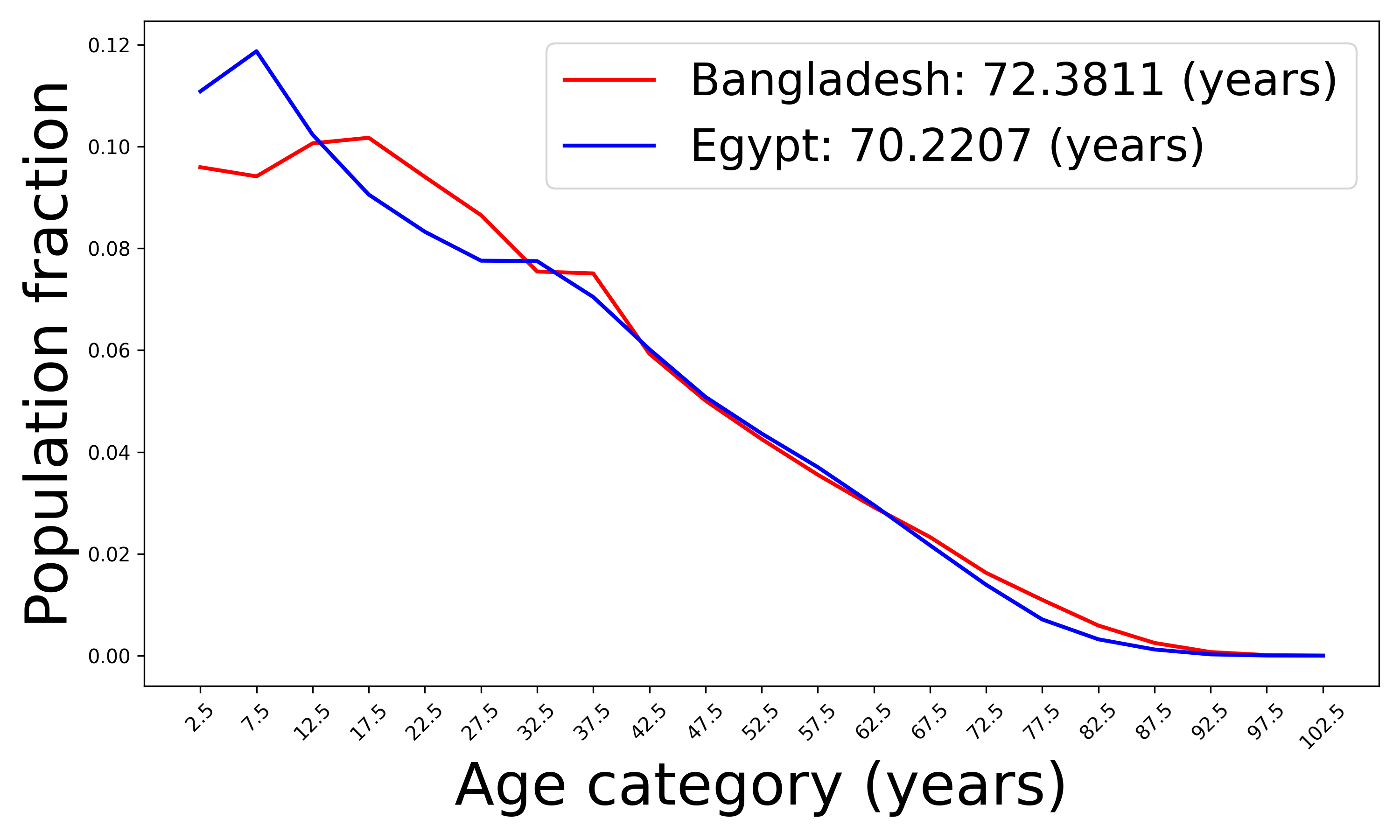}
    \caption{Difference in life expectancy for similar population distribution.}
    \label{fig3:life_structure}
  \end{subfigure}%
  \hspace{1cm}
  \begin{subfigure}{0.3\textwidth}
    \centering
    \includegraphics[width=\linewidth]{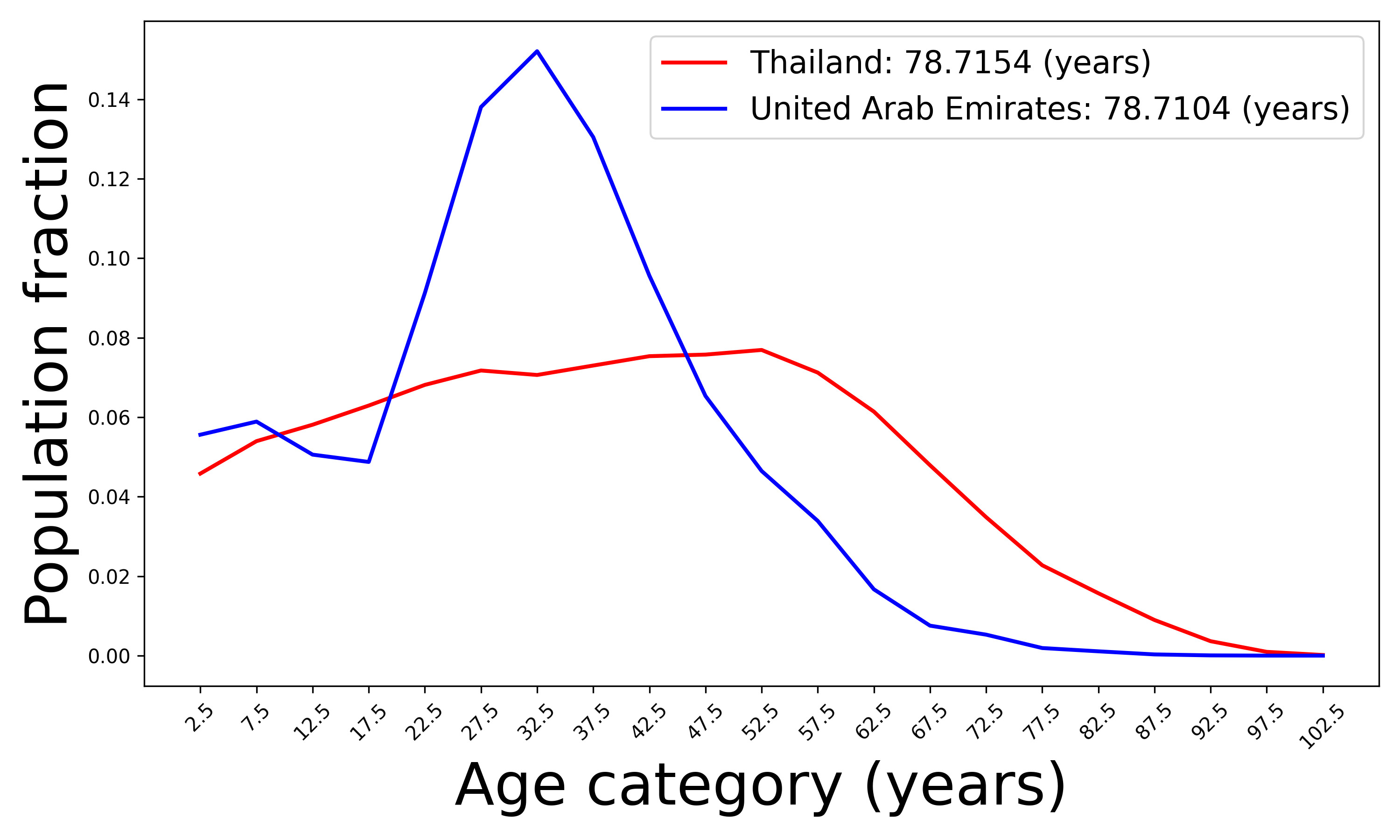}
    \caption{Difference in population distribution for similar life expectancy.}
    \label{fig3:life_index}
  \end{subfigure}
  
  \vspace{0.5cm}
  
  \begin{subfigure}{0.3\textwidth}
    \centering
    \includegraphics[width=\linewidth]{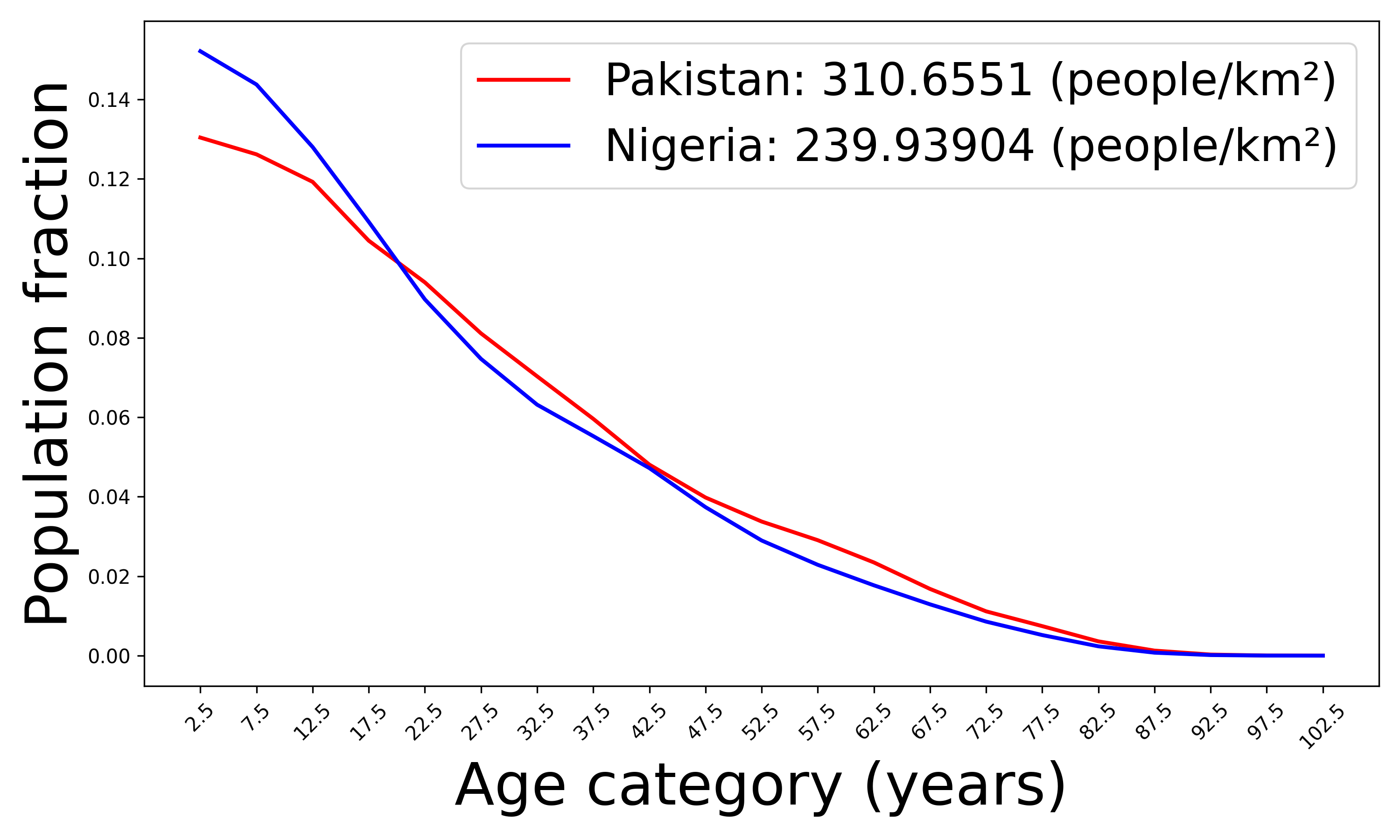}
    \caption{Difference in population density for similar population distribution.}
    \label{fig3:density_structure}
  \end{subfigure}%
  \hspace{1cm}
  \begin{subfigure}{0.3\textwidth}
    \centering
    \includegraphics[width=\linewidth]{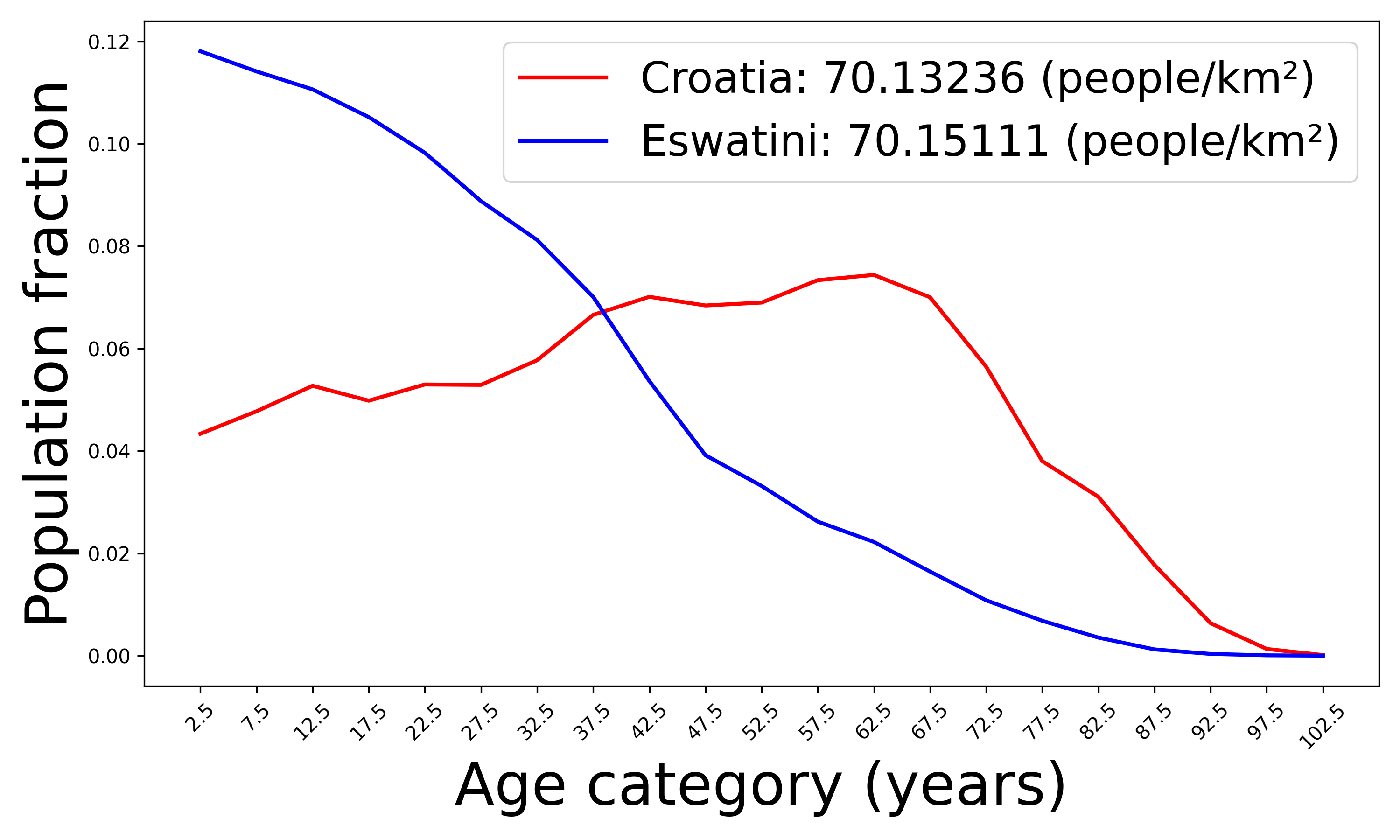}
    \caption{Difference in population distribution for similar population density.}
    \label{fig3:density_index}
  \end{subfigure}
  
  \caption{\small Differences in median age, life expectancy, and population density, together with the corresponding variations in population structure.}
  \label{fig3:all_images}
\end{figure}

Demographic characteristics of the population, median age, life expectancy, and population density have been frequently utilized to study cause specific mortality\cite{D_N._2020,A_Wang_2020}. Research has shown how populational attributes influence the spread and severity of both communicable and non-communicable diseases\cite{H_Ward_1981,R_Nayagam_2016}. High population density has been linked to increased transmission rates of infectious diseases\cite{E_Li_2018}, while aging populations exhibit higher mortality rates for conditions such as cardiovascular diseases and neurodegenerative disorders\cite{I_Ruth_2022}. However, conventional demographic indicators have inherent limitations. Using metrics related to the central tendency of population distribution is insufficient for understanding nuanced demographic influences on mortality patterns. Figures \ref{fig3:all_images} illustrates these discrepancies between demographic indicators and their respective population distributions underscoring the limitations of traditional metrics.

\begin{figure}[htpt]
    \centering
    \begin{subfigure}[b]{0.18\textwidth}
        \centering
        \includegraphics[width=\textwidth]{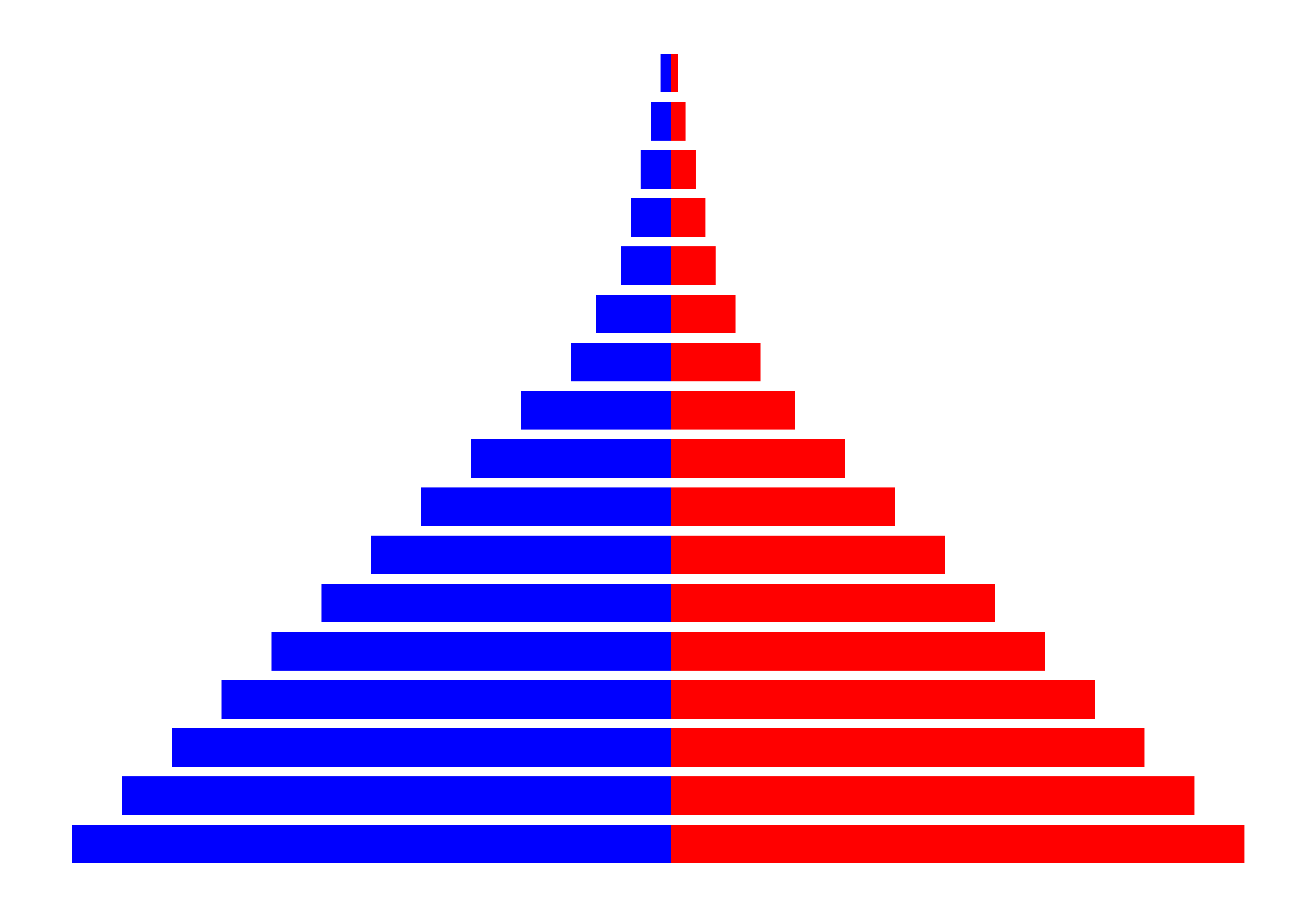}
        \caption{Expansive}
        \label{fig1:image1}
    \end{subfigure}
    \begin{subfigure}[b]{0.18\textwidth}
        \centering
        \includegraphics[width=\textwidth]{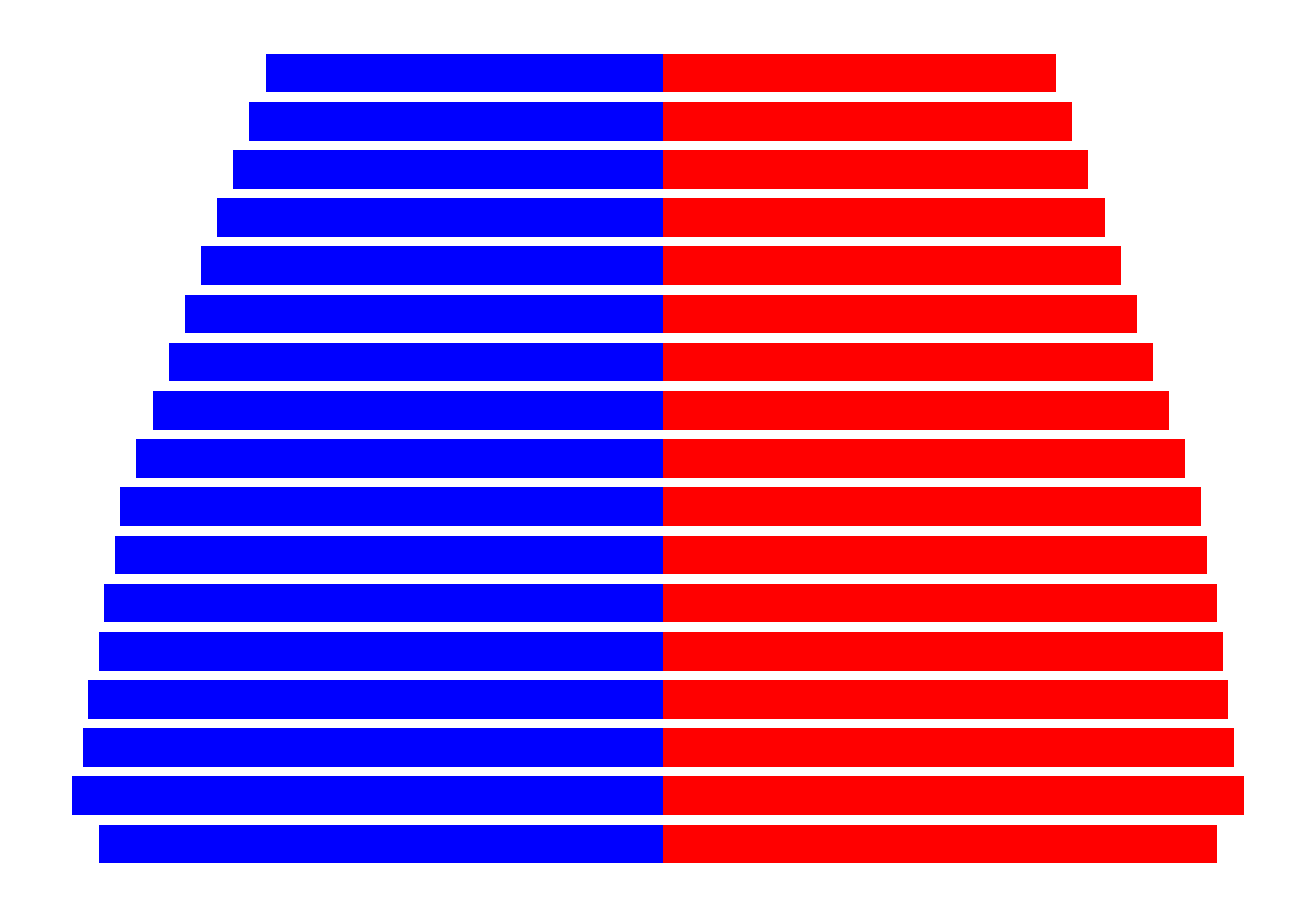}
        \caption{Stationary}
        \label{fig1:image2}
    \end{subfigure}
    \begin{subfigure}[b]{0.18\textwidth}
        \centering
        \includegraphics[width=\textwidth]{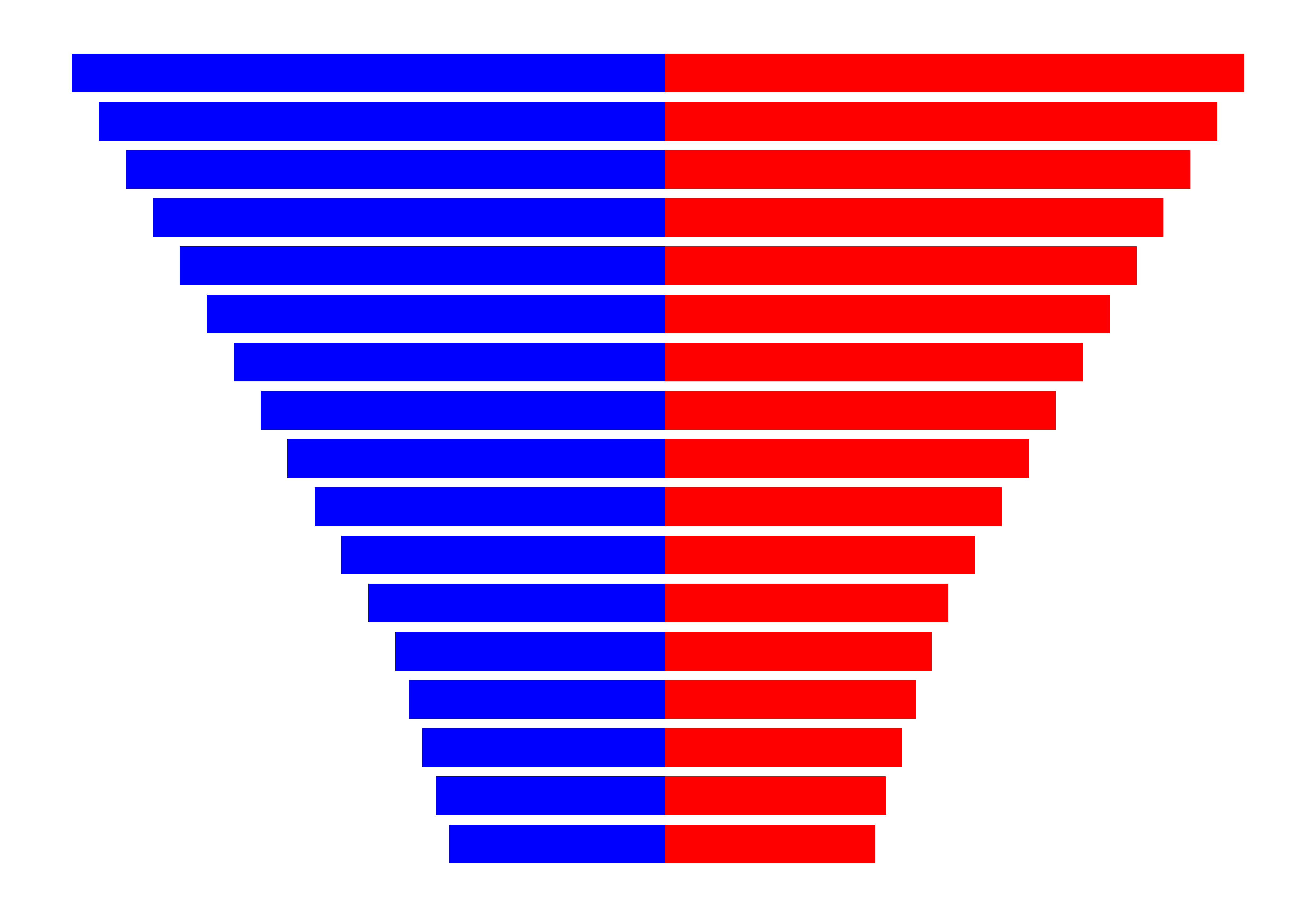}
        \caption{Constrictive}
        \label{fig1:image3}
    \end{subfigure}
    \caption{Prominent population pyramid structures}
    \label{fig1:all_images}
\end{figure}

A more comprehensive approach to understanding demographic structure is through population pyramids which represents the distribution of a population by age and sex. In population pyramids, age is typically divided into five year strata that are stacked vertically in an ascending order. Each strata represents the numbers of both males and females in that age group\cite{B_J._2013}. As shown in Figure \ref{fig1:all_images}, population pyramids are categorized into expansive (progressive), stationary, and constrictive (regressive) forms, reflecting youthful, stable, and aging populations, respectively\cite{Wwjmrd2018}. This unique visualization of population pyramids allows them to capture the entire age-sex spectrum of a country overcoming the limitations of traditional demographic variables. Nevertheless, no methodological approach has yet been proposed in literature to condense these information rich structures into comparable scalar variables. 

The evolution of the population pyramid from an expansive to a constrictive shape is best explained by the landmark Demographic Transition Model (DTM) introduced by Warren Thompson.  Though this transformation is directly driven by birth, death, and migration rates as proposed by the DTM, it is also closely associated with policy changes, economic shifts, natural disasters and, more importantly, epidemiological trends\cite{E_Wang_2016}. Building on the DTM, Abdel Omran’s Epidemiological Transition Model (ETM) posited a direct relationship between transformations in population structures and transitions in cause-specific mortality \cite{T_R._2005}. Therefore, among demographic variables, examining the association of population pyramids with cause specific mortality becomes a priority. However, a thorough literature review revealed that this relationship has been mainly assessed through narrative syntheses\cite{A_Khan_2022} suggesting population pyramids have been underutilized in quantitative epidemiology, limiting their role in mortality studies.

Therefore, this study has two primary objectives, (1) to develop a quantitative method for extracting and condensing age-sex distribution data from population pyramids into a scalar variable, and (2) to employ this variable to investigate the association between disease-specific mortality rates and underlying demographics. 


\section{Methods}
\label{sec:headings}
\subsection{Data Sources and Preprocessing}

Mortality data for 371 diseases and injuries across 180 countries were obtained from the Global Burden of Disease Study 2021 (IHME). Population data were sourced from the United Nations World Population Prospects 2024, with 2021 country specific population pyramids constructed using five year age groups and normalized to reflect gender specific proportions. Causes were organized in a four tiered hierarchy, and for our analysis, fatal outcomes up to Level 3 were considered.

\subsection{Constructing the PoPDivergence}

Population pyramids are multidimensional representations of population structure. However, a scalar metric is needed to correlate them with mortality rates. We propose PoPDivergence as shown in \ref{eq1}, a Kullback Leibler (KL) divergence based measure to quantify the structural deviation of a population pyramid from an optimized reference population.
\begin{equation}
    PoPDivergence(ref:Q) = \sum_{i \in I}P_{(i)}log\left(\frac{P_{(i)}}{Q_{(i)}}\right) 
    \label{eq1}
\end{equation}
Here, $P$ and $Q$ denote population distributions, with $P_{(i)}$ and $Q_{(i)}$ representing the proportion of the population in the $i^{th}$ age category, and $Q$ serving as the optimized reference population.

\begin{figure}[htbp]
    \centering
    \hspace*{-2cm}
    \begin{tabular}{@{}cccccc@{}}
        {\footnotesize PopDivergence} & 0 & 0.009 & 0.186 & 0.643 & 0.664 \\[1ex]
         & \includegraphics[width=0.18\textwidth]{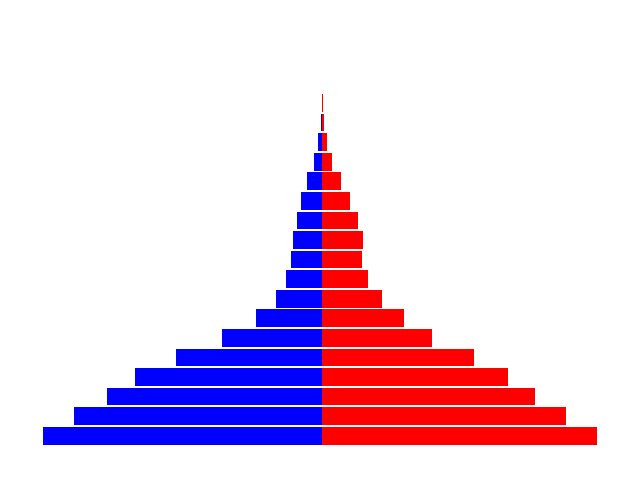} 
         & \includegraphics[width=0.18\textwidth]{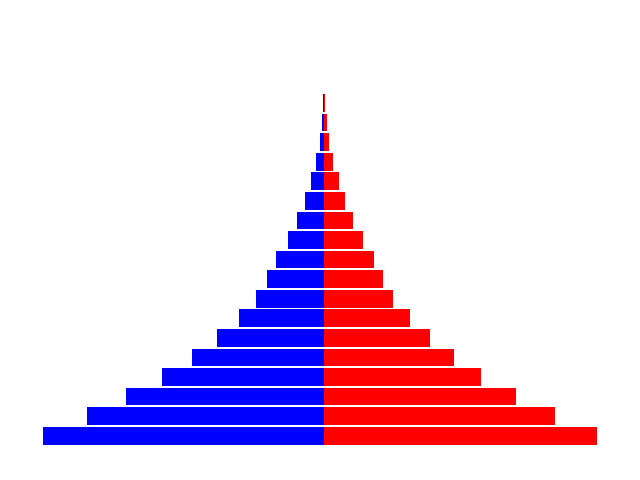} 
         & \includegraphics[width=0.18\textwidth]{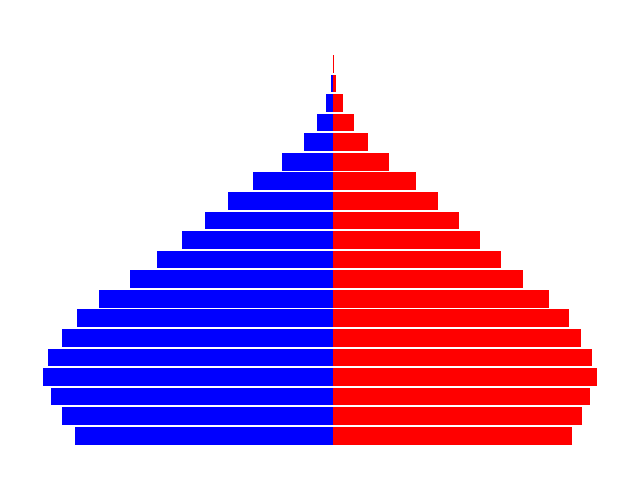} 
         & \includegraphics[width=0.18\textwidth]{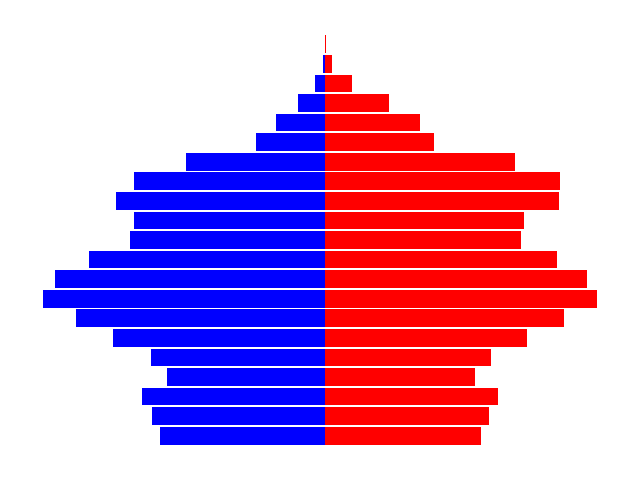} 
         & \includegraphics[width=0.18\textwidth]{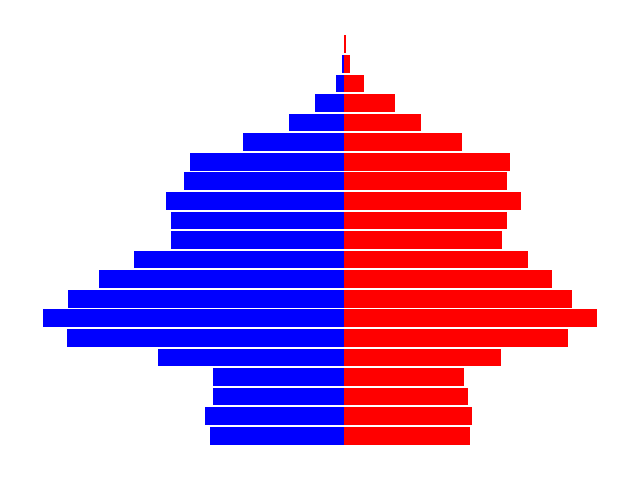} \\[1ex]
         & Central African Rep. & Chad & India & Poland & Malta \\
         \\[2ex]
        {\footnotesize PopDivergence} & 0 & 0.026 & 0.079 & 0.320 & 0.518 \\[1ex]
         & \includegraphics[width=0.18\textwidth]{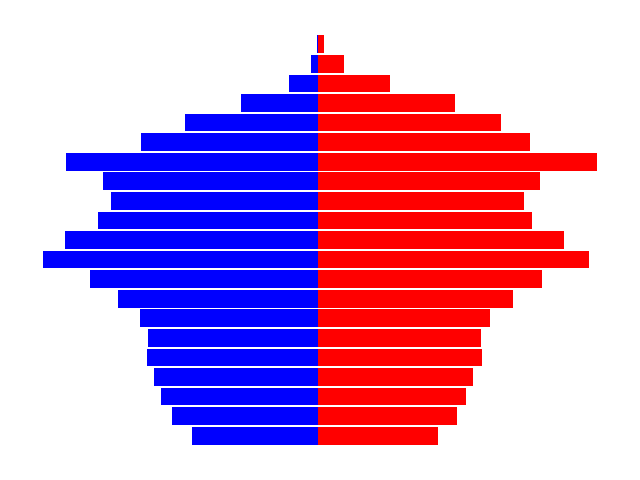} 
         & \includegraphics[width=0.18\textwidth]{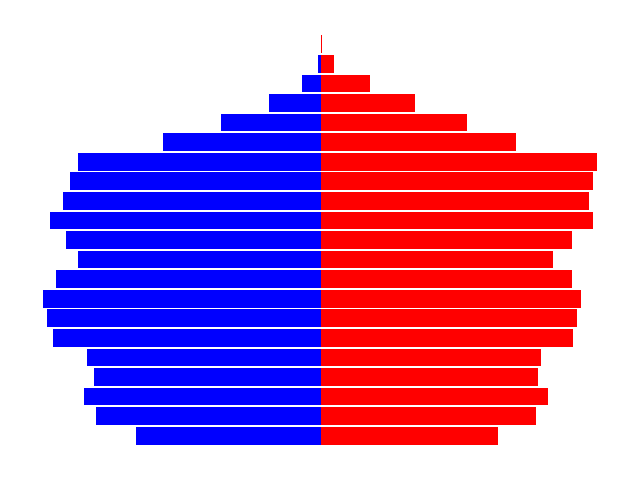} 
         & \includegraphics[width=0.18\textwidth]{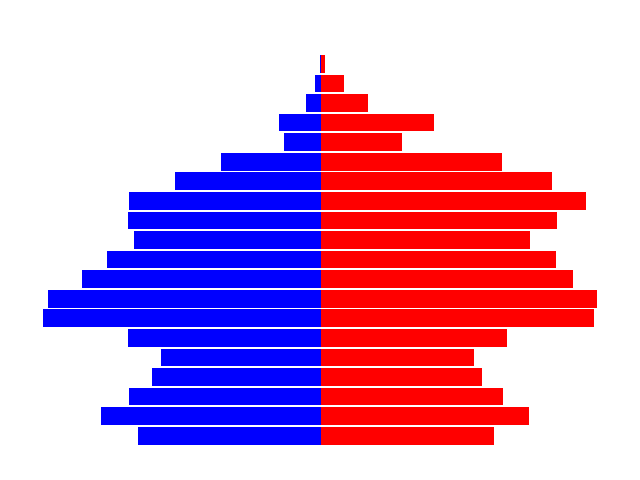} 
         & \includegraphics[width=0.18\textwidth]{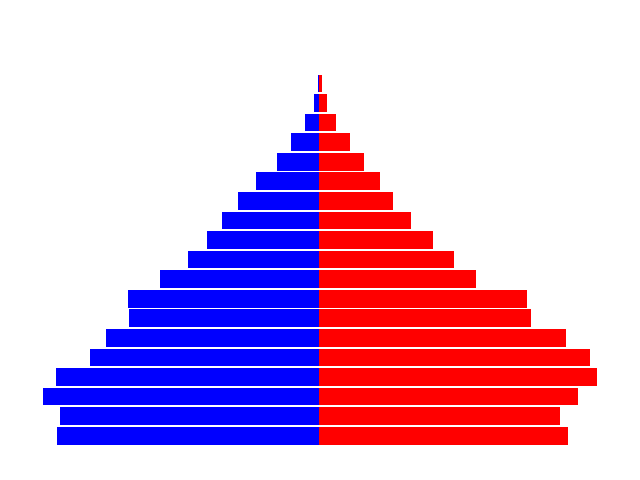} 
         & \includegraphics[width=0.18\textwidth]{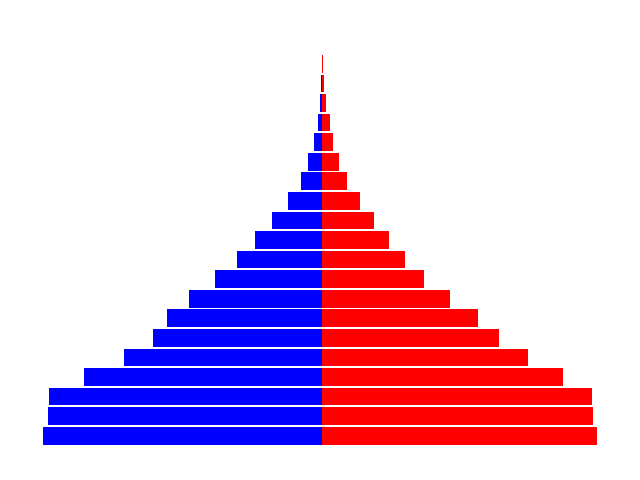} \\[1ex]
         & Japan & Finland & Russia & Bangladesh & Kenya \\
    \end{tabular}
    \caption{Comparison of PoPDivergence values across countries. Top row: Central African Republic as reference, showing transition from progressive to regressive population pyramids. Bottom row: Japan as reference, indicating movement toward progressive structures.}
    \label{fig:pop_div_combined}
\end{figure}

This measures the information theoretic distance between demographic structures. Figures \ref{fig:pop_div_combined} illustrates PoPDivergence using the Central African Republic (CAR) and Japan as reference populations, deliberately chosen to represent the extremes of the demographic transition. A zero value indicates identical distributions. Contextual interpretation depends on the reference population pyramid.

\subsection{Reference Country Tuning}
Since, the PoPDivergence values depend on a pre selected reference country, to maximize correlation with cause specific mortality rates (expressed as the natural log of deaths per million), we optimize the reference selection using the objective function \ref{eq2},

\begin{equation}
    \underset{\omega \in \Omega}{\arg{max}} \;\;Cor \left[\ln{S}, PoPDivergence(\omega)\right]
    \label{eq2}
\end{equation}

where $\Omega$ is the set of countries, and $S$ is the mortality rates. Given the absence of a closed form solution, brute force search identifies the reference pyramid with the strongest correlation to mortality. This method is what we introduce as reference country tuning, which enables the identification of the optimized population demographic(out of 180 countries), which is most strongly associated for a given disease’s mortality rate.

\subsection{Constructing the PoPStat}
PoPStat quantifies the association between cause specific mortality and the structure of population pyramids. For each cause, a reference country is first selected through reference tuning. Then, the PoPDivergence from this reference is calculated for every country (stored in vector X), while the natural logarithms of the corresponding cause specific death rates are stored in vector Y. PoPStat is then defined as the Pearson correlation coefficient between X and Y.

\begin{equation}
    PoPStat(ref) = \frac{
    \sum_{i=1}^n (x_i-\Bar{x})(y_i-\Bar{y})
    }{
    \sqrt{
    \sum_{i=1}^n (x_i - \Bar{x})^2 (y_i-\Bar{y})^2
    }
    }
    \label{eq3}
\end{equation}
Here, $x_i$ denotes the PoPDivergence of the $i^{\text{th}}$ country, and $y_i$ is the natural logarithm of its cause specific mortality rate. The terms $\bar{x}$ and $\bar{y}$ represent the means of vectors $X$ and $Y$, respectively.

In this study, we will consistently present Popstat in the format (PopStat value, reference country, p-value).

\subsection{Comparison of PoPStat with other indicators}

To benchmark PoPStat, its association with cause specific mortality was compared with traditional demographic indices, life expectancy, median age, and population density, as well as socioeconomic measures such as GDP per capita and the Human Development Index (HDI). Pearson correlation coefficients between each indicator and the cause specific death rates were calculated to evaluate their relative predictive strengths. 

\section{Results}
\begin{figure}[htbp]
    \centering
    \begin{subfigure}[t]{0.3\textwidth}
        \centering
        \includegraphics[width=\textwidth]{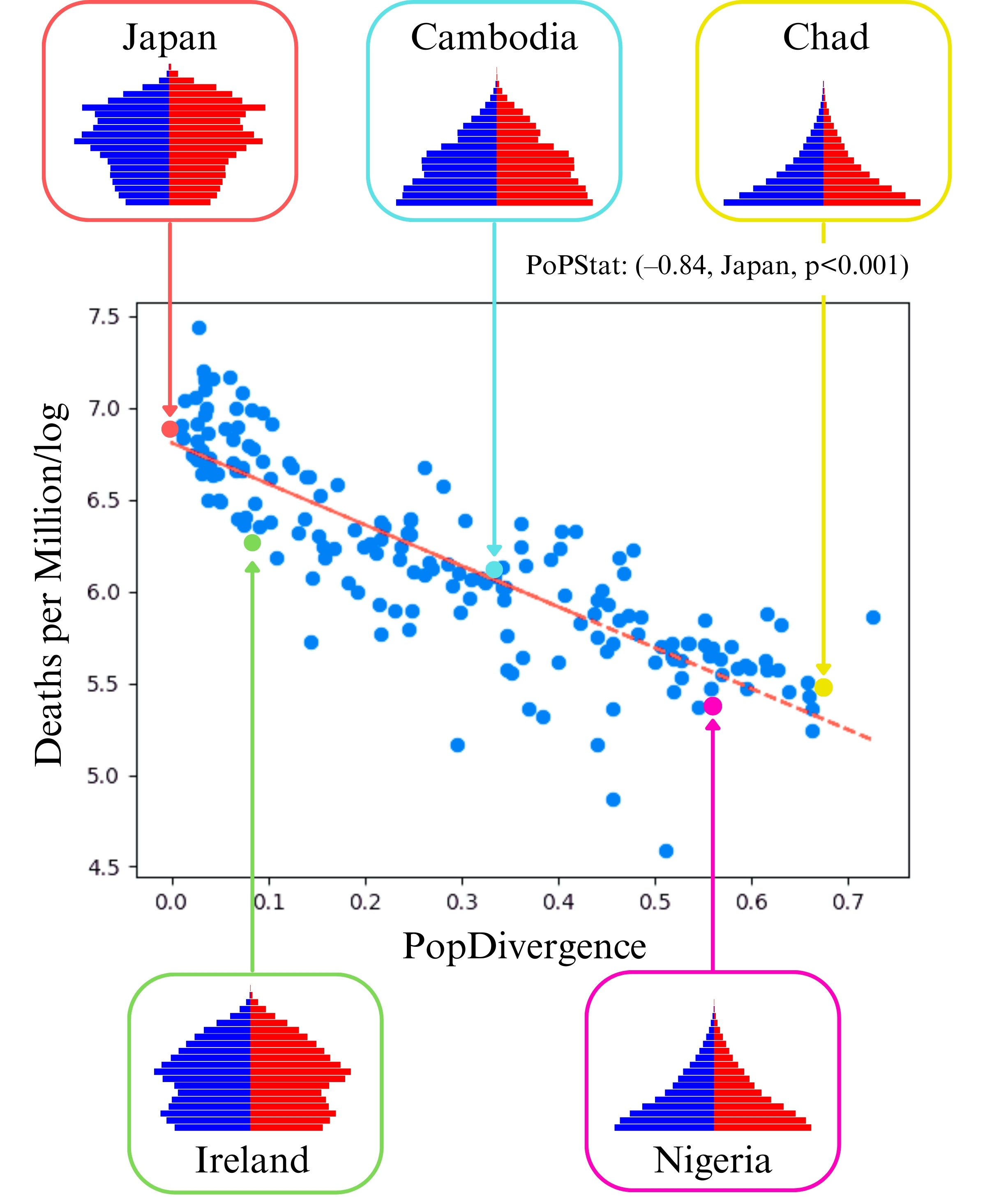}
        \caption{Non-Communicable Diseases}
        \label{fig7:non_communicable}
    \end{subfigure}
    \hfill
    \begin{subfigure}[t]{0.3\textwidth}
        \centering
        \includegraphics[width=\textwidth]{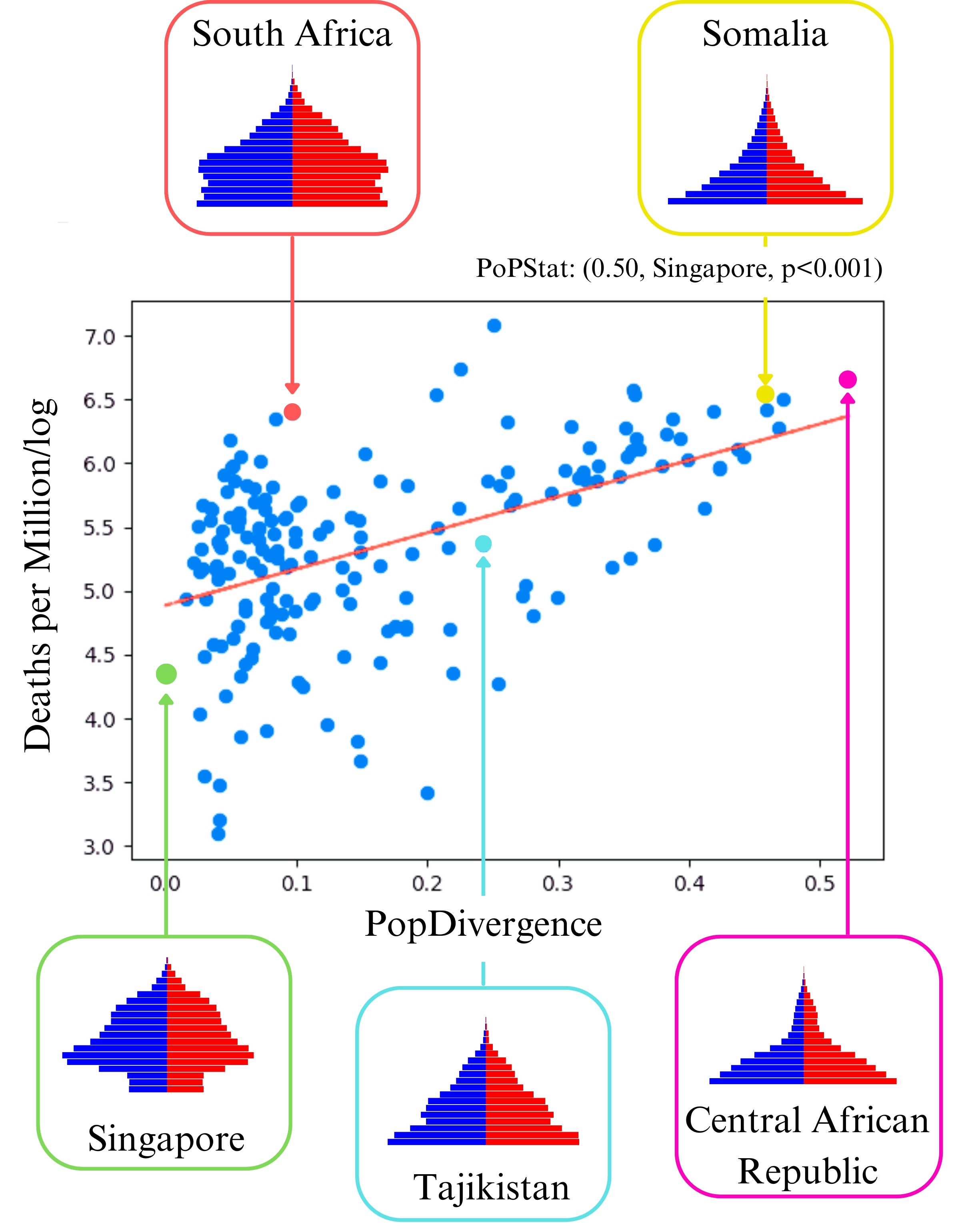}
        \caption{Communicable, Maternal, Neonatal, and Nutritional Diseases}
        \label{fig7:cmnn}
    \end{subfigure}
    \hfill
    \begin{subfigure}[t]{0.3\textwidth}
        \centering
        \includegraphics[width=\textwidth]{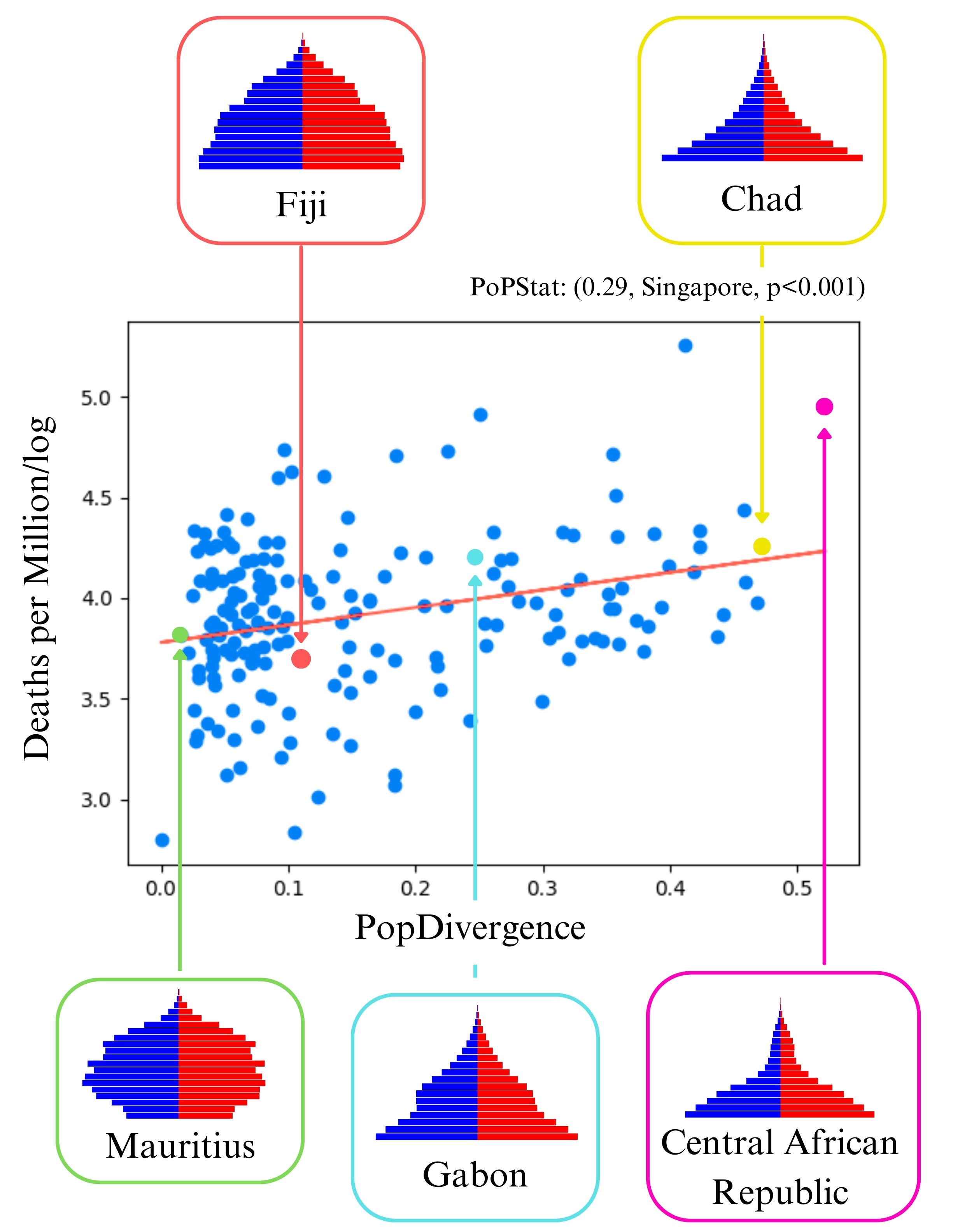}
        \caption{Injuries}
        \label{fig7:injuries}
    \end{subfigure}
    
    \vspace{1em} 
    
    \begin{subfigure}[t]{0.3\textwidth}
        \centering
        \includegraphics[width=\textwidth]{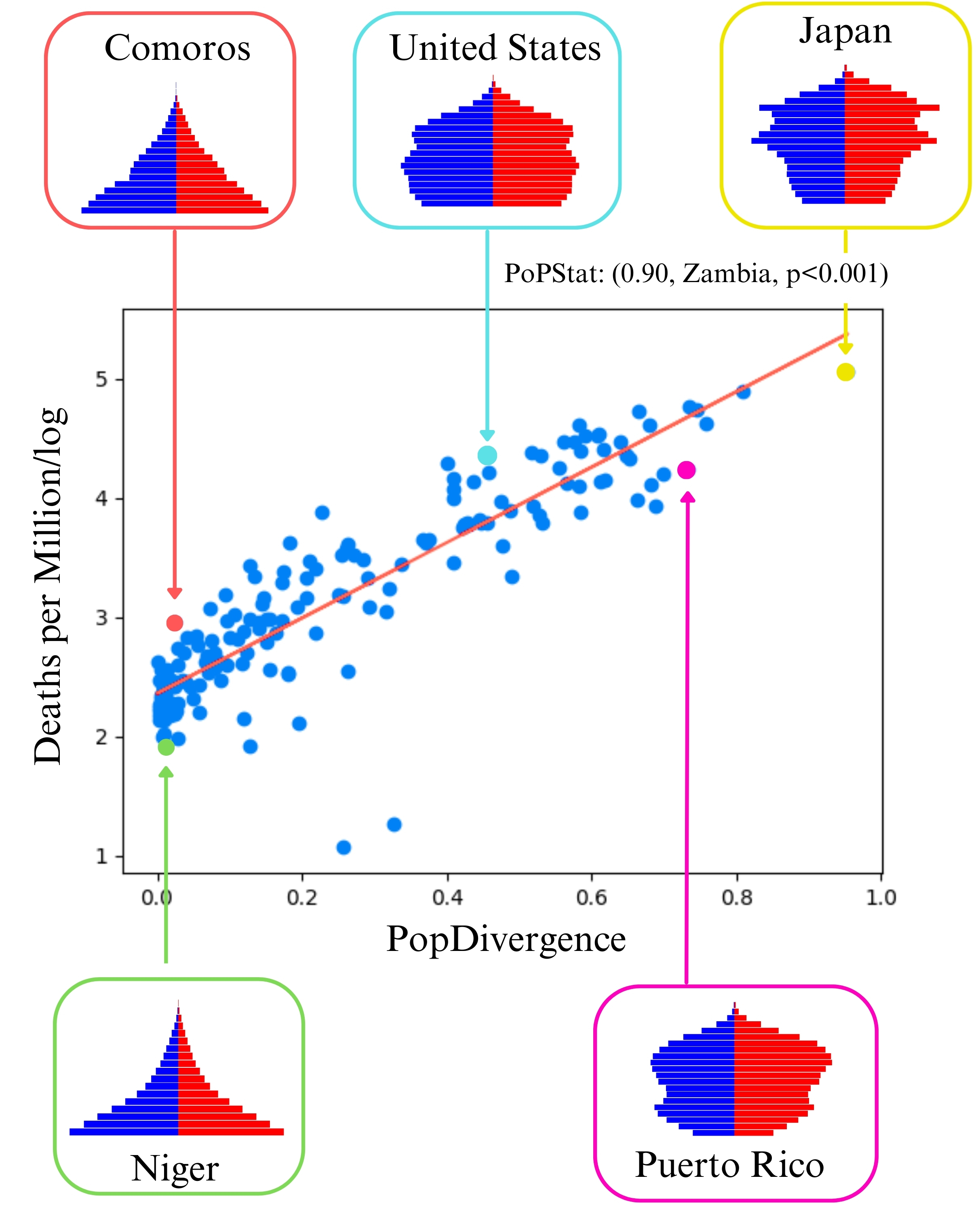}
        \caption{Neurological Disorders}
        \label{fig7:neurological}
    \end{subfigure}
    \hfill
    \begin{subfigure}[t]{0.3\textwidth}
        \centering
        \includegraphics[width=\textwidth]{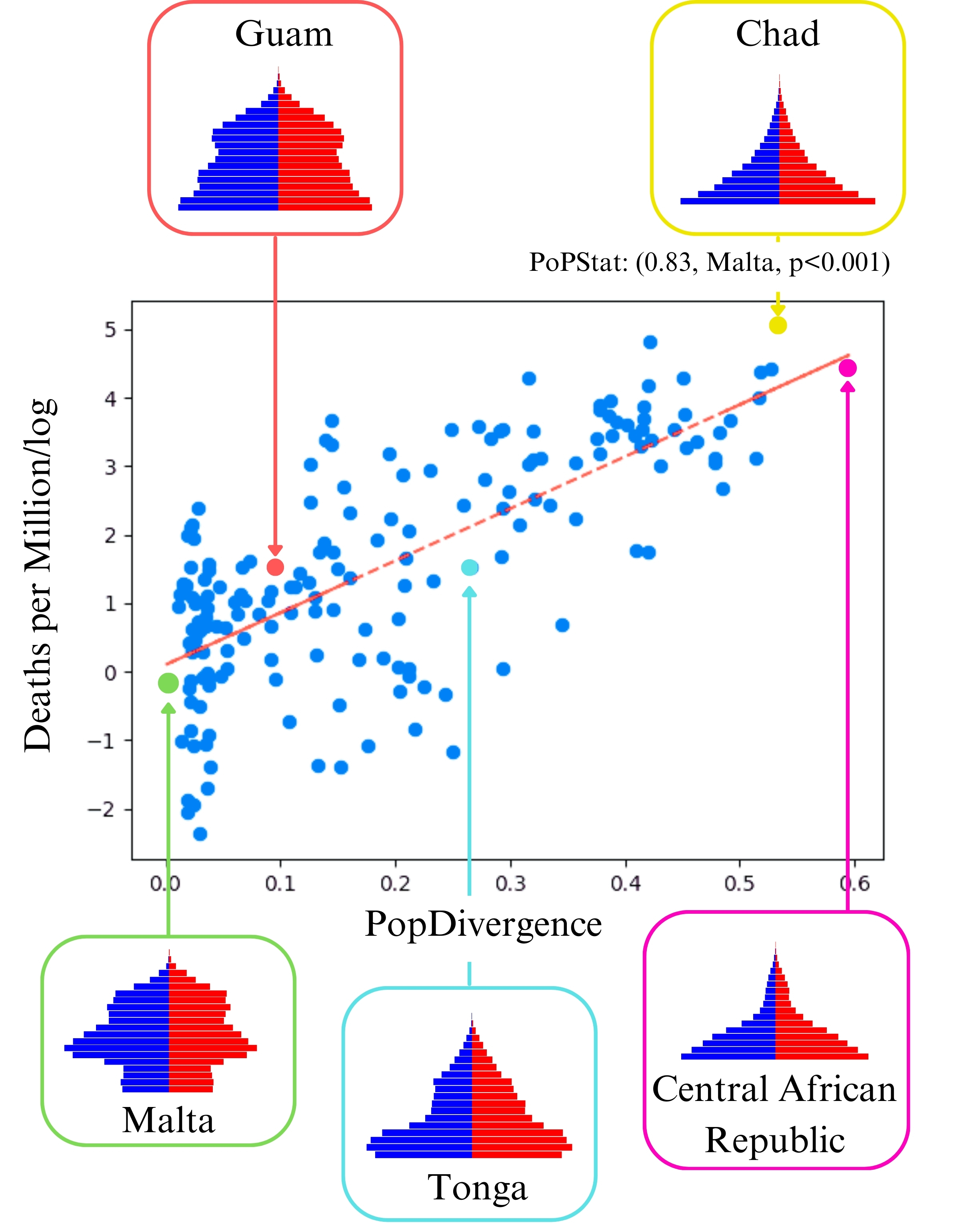}
        \caption{Neglected Tropical Diseases}
        \label{fig7:ntd}
    \end{subfigure}
    \hfill
    \begin{subfigure}[t]{0.3\textwidth}
        \centering
        \includegraphics[width=\textwidth]{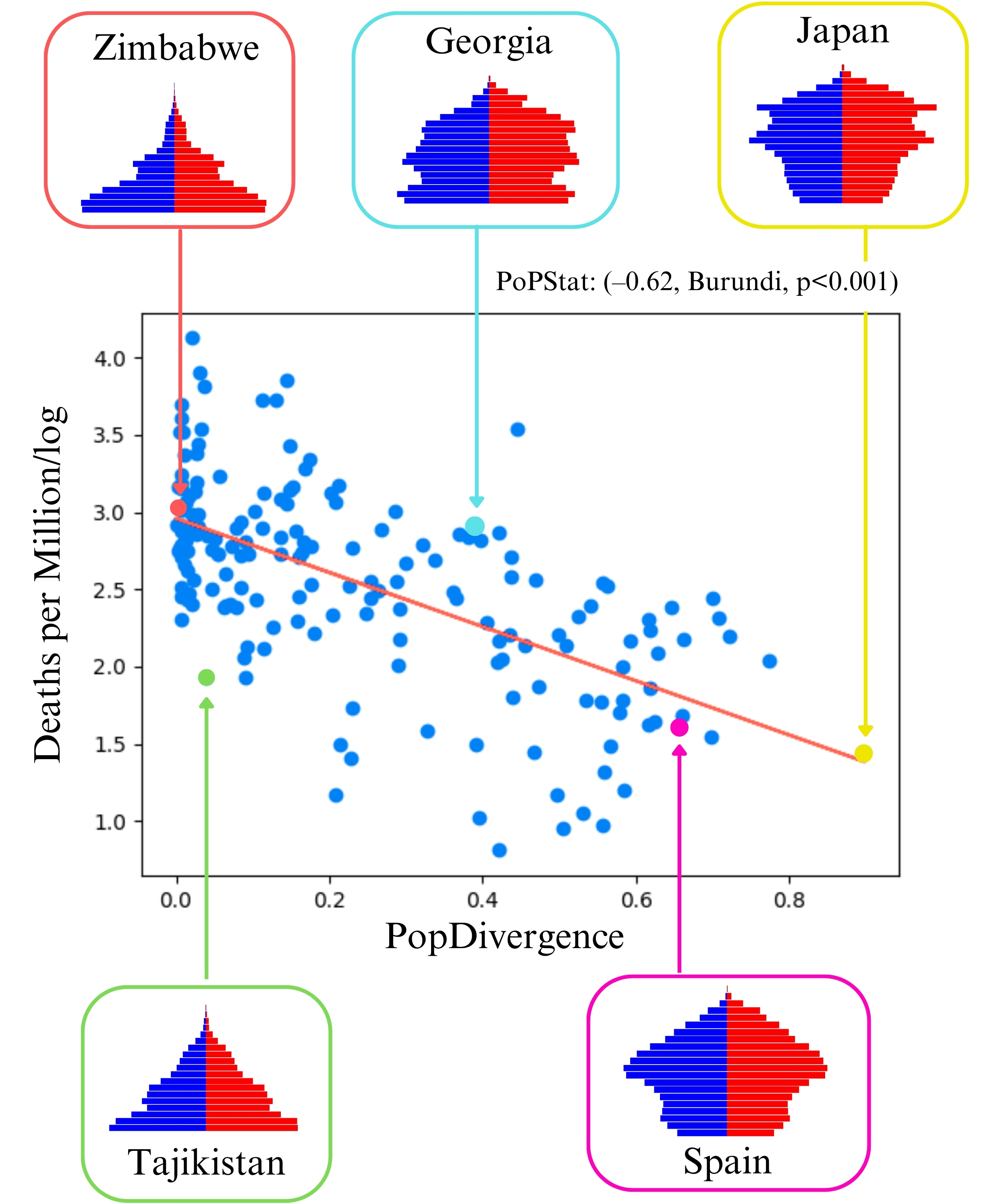}
        \caption{Transport Injuries}
        \label{fig7:transport}
    \end{subfigure}
    
    \vspace{1em} 
    
    \begin{subfigure}[t]{0.3\textwidth}
        \centering
        \includegraphics[width=\textwidth]{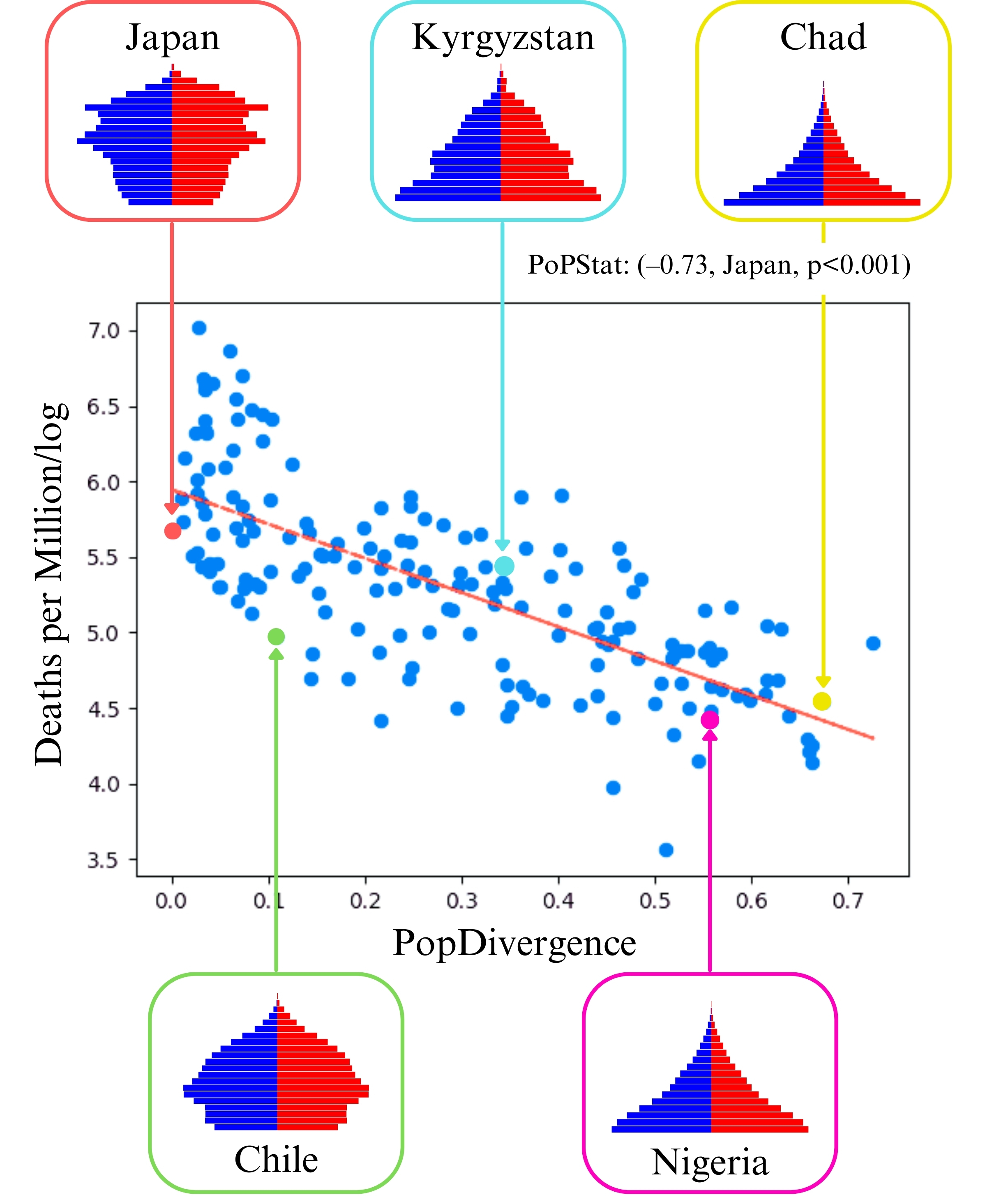}
        \caption{Cardiovascular Diseases}
        \label{fig7:cardiovascular}
    \end{subfigure}
    \hspace{0.5em}
    \begin{subfigure}[t]{0.3\textwidth}
        \centering
        \includegraphics[width=\textwidth]{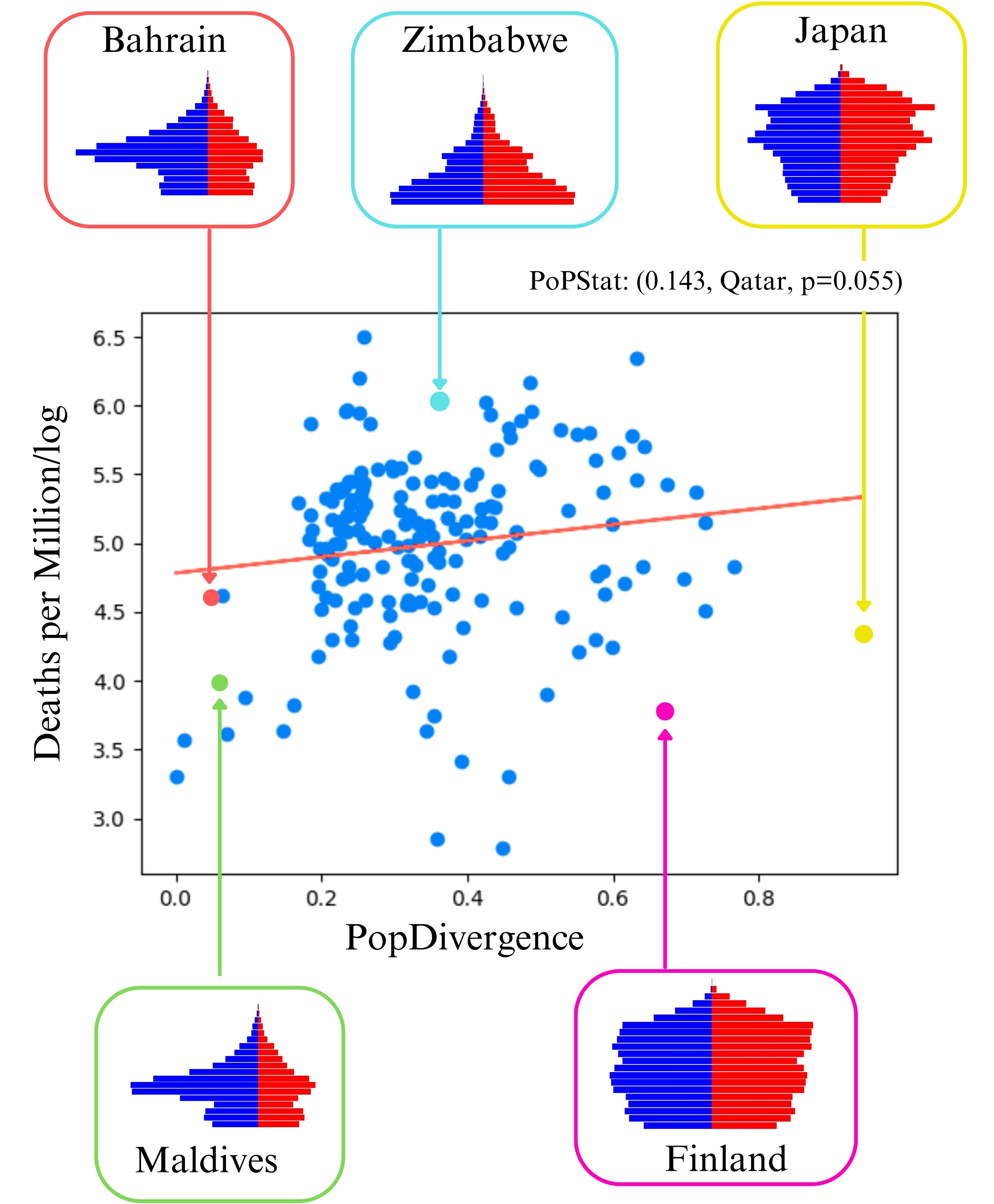}
        \caption{Respiratory Infections and Tuberculosis}
        \label{fig7:respiratory}
    \end{subfigure}
    
    \caption{PoPDiveregence vs Disease Mortality for Selected High Impact Disease Categories}
    \label{fig7:combined_disease_categories}
\end{figure}

PoPStat values were calculated for all disease categories up to Level 3 are summarized in Table \ref{tab:float_table_1} and \ref{tab:float_table_2}. Figure \ref{fig7:combined_disease_categories}, shows variations of PoPDivergence and mortality rates for selected disease categories, where population pyramids of randomly selected countries are shown for each plot to illustrate the variation of PoPDivergence.

At Level 1, among the three aggregated disease categories, non communicable diseases (NCD) exhibited the highest association with overall mortality, with a PoPStat of (–0.84, Japan, p<0.001), explaining 81.4\% of the mortality variation. As shown in Figure \ref{fig7:non_communicable} countries with constrictive and more symmetrically shaped population pyramids such as Japan, Italy, Portugal, and Greece had higher mortality due to NCDs. In contrast, the aggregate of communicable, maternal, neonatal, and nutritional diseases (CMNN) showed a moderate association with a PoPStat of (0.50, Singapore, p<0.001), accounting for 25.4\% of the variation as shown in Figure \ref{fig7:cmnn}. Countries with more symmetrical and constrictive pyramids like Singapore, Mauritius, Chile, and Saint Lucia saw lower CMNN death rates. Injuries, as visualized in Figure \ref{fig7:injuries}, exhibited the weakest association, with a PoPStat of (0.29, Singapore, p<0.001), explaining only 8.4\% of the variation in injury‐related mortality. Countries like Singapore, Mauritius, Chile, and Saint Lucia with constrictive pyramids that were largely symmetrical had lower injury related mortality.

Within the NCD group at Level 2, neurological diseases recorded the highest PoPStat of (0.90, Zambia, p<0.001), explaining 81.4\% of the mortality variation attributed to neurological causes as depicted in Figure \ref{fig7:neurological}. Countries like Burkina Faso, Cameroon, and Yemen, with expansive, predominantly symmetrical population pyramids, had a lower burden of deaths due to neurological diseases. Within the CMNN group at Level 2, under communicable diseases, neglected tropical diseases (NTD) had the highest PoPStat of (0.83, Malta, p<0.001), which explained 69.7\% of the NTD‐related mortality variation, as visualized in Figure \ref{fig7:ntd}.  Countries with symmetrical, constrictive population pyramids such as Malta, Poland, Cyprus, and Slovakia experienced lower mortality from NTDs. Under maternal, neonatal, and nutritional diseases, maternal and neonatal conditions exhibited the highest PoPStat(0.91, Central African Republic, p<0.001), explaining 83.6\% of the variance in maternal and neonatal mortality. Countries such as Niger, Somalia, and Chad which shared expansive population pyramids
saw higher death rates. Within the injuries category, transport injuries (TI), as shown in Figure \ref{fig7:transport} indicated a moderate association with a PoPStat of (–0.62, Burundi, p<0.001), explaining 38.5\% of the mortality variation. Countries like Mali, Burkina Faso, and Mozambique showed similar symmetrically expansive pyramids and higher mortality.

At Level 2, focusing on the most severe subcategories, cardiovascular diseases (CVD) as the leading contributor among NCDs, exhibits a strong association of (–0.73, Japan, p<0.001), explaining 53.4\% of the mortality variation. Figure \ref{fig7:cardiovascular}, illustrates how countries with symmetrical and constrictive population
pyramids identified before had more death rates due to cardiovascular diseases. Conversely, within the CMNN category, respiratory infections and tuberculosis showed a weak association, with a PoPStat of (0.143, Qatar, p=0.055) that explained only 2\% of the mortality variation. As seen in \ref{fig7:respiratory}, asymmetrically constrictive pyramids like Qatar, UAE, Bahrain, and Maldives had a lower burden of deaths.

\begin{figure}[htbp]
    \centering
    \begin{subfigure}[t]{0.33\textwidth}
        \centering
        \includegraphics[width=\textwidth]{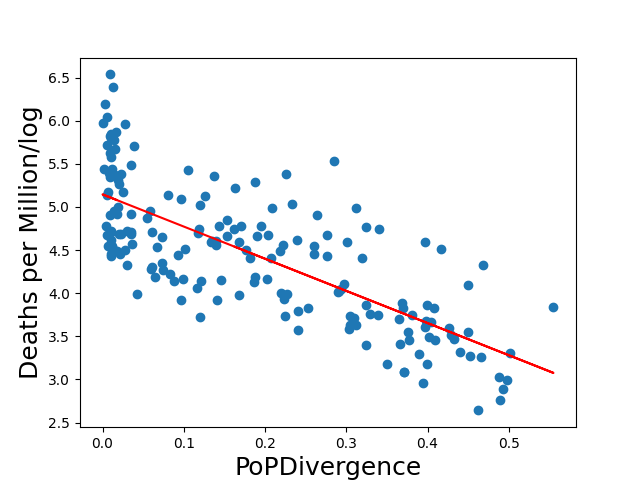}
        \caption{PoPStat: (\textbf{$-0.750$}, Latvia, $p:<0.001$)}
        \label{fig6:first_image}
    \end{subfigure}
    \begin{subfigure}[t]{0.33\textwidth}
        \centering
        \includegraphics[width=\textwidth]{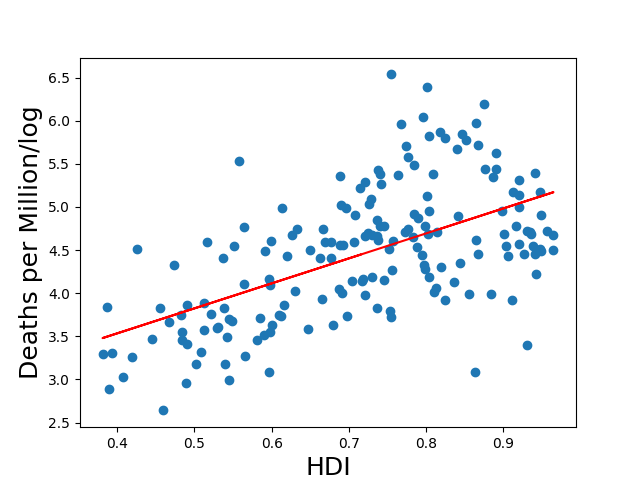}
        \caption{HDI: correlation coefficient:$r=0.575$}
        \label{fig6:second_image}
    \end{subfigure}
    \begin{subfigure}[t]{0.33\textwidth}
        \centering
        \includegraphics[width=\textwidth]{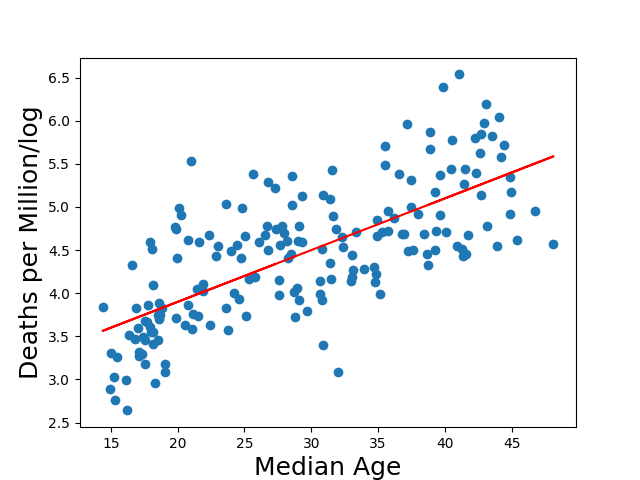}
        \caption{Median Age: $r=0.708$}
        \label{fig6:third_image}
    \end{subfigure}

    \begin{subfigure}[t]{0.33\textwidth} 
        \centering
        \includegraphics[width=\textwidth]{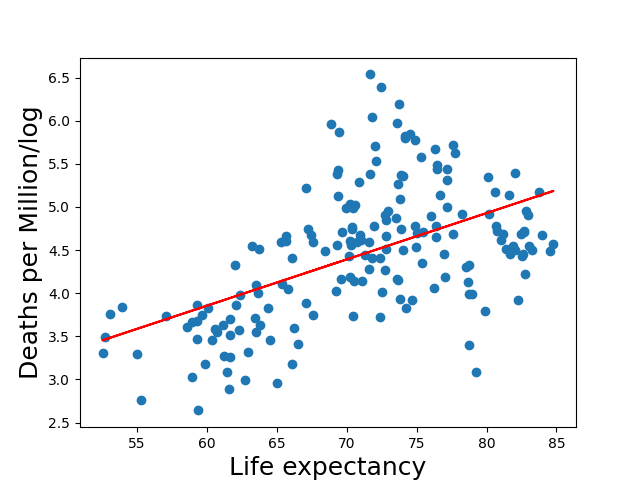}
        \caption{Life Expectancy: $r=0.528$}
        \label{fig6:plot_below}
    \end{subfigure}
    \begin{subfigure}[t]{0.33\textwidth} 
        \centering
        \includegraphics[width=\textwidth]{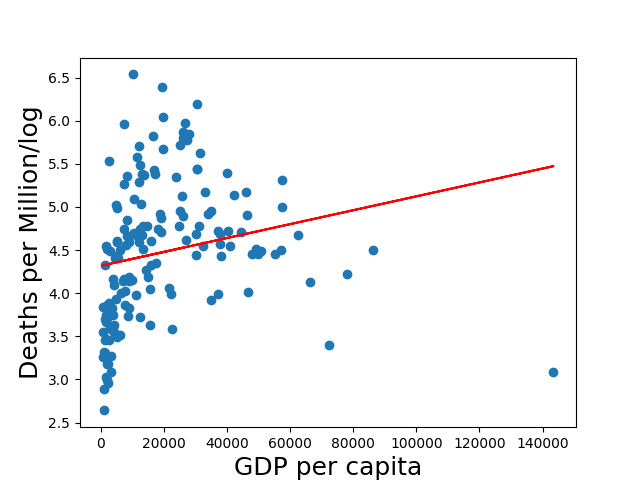}
        \caption{GDP per capita: $r=0.206$}
        \label{fig6:plot_below2}
    \end{subfigure}
    \begin{subfigure}[t]{0.33\textwidth} 
        \centering
        \includegraphics[width=\textwidth]{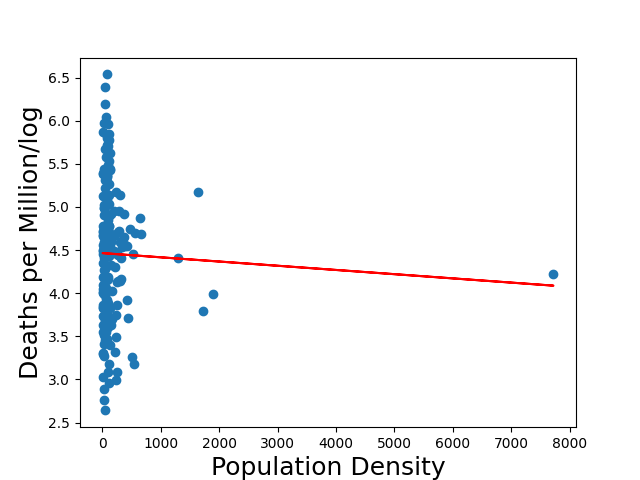}
        \caption{Population Density: $r=-0.039$}
        \label{fig6:plot_below3}
    \end{subfigure}
    
    \caption{PoPStat of Ischemic heart diseases against correlations with other indicators}
    \label{fig6:main_figure}
\end{figure}

Diving further on the most severe diseases in the CVD subgroup at Level 3, ischemic heart disease (IHD) demonstrated a robust association with a PoPStat of (–0.75, Latvia, p<0.001), explaining 56.2\% of IHD related mortality. We use IHD as an example to calculate the correlations of other well recognized demographic and socio economic indicators with mortality rates and to compare their performance with PoPStat, as seen in \ref{fig6:first_image}.

\raggedbottom

\begin{table}[htbp]
\centering
\caption{PoPStat and Reference Populations for Level 1 Diseases and Selected Level 2 Diseases}
\label{tab:float_table_1}
\footnotesize  
\renewcommand{\arraystretch}{1.2}
\rowcolors{2}{lightgray!30}{white}
\begin{tabular}{%
    >{\raggedright\arraybackslash}p{3.7cm}
    >{\centering\arraybackslash}p{0.9cm}
    >{\centering\arraybackslash}p{1.2cm}
    >{\centering\arraybackslash}p{1cm}
    >{\centering\arraybackslash}p{2.1cm}
    >{\centering\arraybackslash}p{1.0cm}
    >{\centering\arraybackslash}p{1.4cm}
    >{\centering\arraybackslash}p{1.4cm}
    >{\centering\arraybackslash}p{1.4cm}}
\toprule
\rowcolor{gray!30} \textbf{Disease Category} & \multicolumn{3}{c}{\textbf{PoPStat}} & \textbf{CI}& \textbf{PoPStat} & \multicolumn{3}{c}{\textbf{Closest countries to the reference}} \\
\rowcolor{gray!30} & {r} & {Reference country} & {p} &  &\textbf{ r$^2$} & {First} & {Second} & {Third} \\
\midrule 
\multicolumn{9}{l}{\textbf{Level 1 Disease Categories}} \\
Non-communicable diseases & -0.846 & Japan & < 0.001 & (-0.883, -0.799) & 0.716 & Italy & Portugal & Greece \\
Communicable , maternal ,  neonatal,  and nutritional diseases & 0.504 & Singapore & < 0.001 & (0.387, 0.605) & 0.254 & Mauritius & Chile & Saint Lucia \\
Injuries & 0.291 & Singapore & < 0.001 & (0.151, 0.419) & 0.084 & Mauritius & Chile & Saint Lucia \\ 
\multicolumn{9}{l}{\textbf{Level 2 Disease Categories}} \\
\multicolumn{9}{l}{\textbf{Non Communicable Diseases}} \\
Neurological disorders & 0.902 & Zambia & < 0.001 &  (0.871,  0.926)  & 0.814 & Burkina Faso & Cameroon & Yemen\\
Neoplasms & -0.897 & Japan & < 0.001 &  (-0.922,  -0.864)  & 0.804 & Italy & Portugal & Greece\\
Cardiovascular diseases & -0.731 & Japan & < 0.001 &  (-0.792,  -0.655)  & 0.534 & Italy & Portugal & Greece\\
Mental disorders & -0.7 & Japan & < 0.001 &  (-0.767,  -0.617)  & 0.49 & Italy & Portugal & Greece\\
Musculoskeletal disorders & -0.646 & Japan & < 0.001 &  (-0.724,  -0.552)  & 0.418 & Italy & Portugal & Greece\\
Chronic respiratory diseases & -0.53 & France & < 0.001 &  (-0.627,  -0.416)  & 0.281 & Belgium & United Kingdom & Finland\\
Substance use disorders & -0.533 & Japan & < 0.001 &  (-0.63,  -0.419)  & 0.284 & Italy & Portugal & Greece\\
Digestive diseases & 0.499 & Qatar & < 0.001 &  (0.381,  0.601)  & 0.249 & United Arab Emirates & Bahrain & Maldives\\
Other non-communicable diseases & 0.487 & Singapore & < 0.001 &  (0.367,  0.591)  & 0.237 & Mauritius & Chile & Saint Lucia\\
Diabetes and kidney diseases & -0.408 & Argentina & < 0.001 &  (-0.523,  -0.278)  & 0.166 & Sri Lanka & Costa Rica & Saint Vincent and the Grenadines\\
\multicolumn{9}{l}{\textbf{Communicable Diseases}} \\
Neglected tropical diseases and malaria & 0.835 & Malta & < 0.001 & (0.785,  0.874) & 0.697 & Poland & Cyprus & Slovakia\\
Other infectious diseases & 0.827 & Malta & < 0.001 & (0.775,  0.869) & 0.685 & Poland & Cyprus & Slovakia\\
Enteric infections & 0.735 & Malta & < 0.001 & (0.66,  0.796) & 0.54 & Poland & Cyprus & Slovakia\\
HIV/AIDS and sexually transmitted infections & -0.644 & Central African Republic & < 0.001 & (-0.722,  -0.55) & 0.415 & Niger & Somalia & Chad\\
Respiratory infections and tuberculosis & 0.143 & Qatar & 0.055 & (-0.003,  0.283) & 0.02 & United Arab Emirates & Bahrain & Maldives\\
\multicolumn{9}{l}{\textbf{ Maternal ,
Neonatal, and Nutritional
Diseases}} \\
Maternal and neonatal disorders & -0.915 & Central African Republic & < 0.001 & (-0.936,  -0.887) & 0.836 & Niger & Somalia & Chad \\
Nutritional deficiencies & 0.569 & Cyprus & < 0.001 & (0.46,  0.66) & 0.324 & Luxembourg & Chile & Iceland \\
\multicolumn{9}{l}{\textbf{Injuries}} \\
Transport injuries & -0.621 & Burundi & < 0.001 & (-0.703,  -0.522) & 0.385 & Mali & Burkina Faso & Mozambique\\
\bottomrule
\end{tabular}
\end{table}

\begin{table}[httbp]
\centering
\caption{PoPStat and Reference Populations for Selected Level 3 Diseases}
\label{tab:float_table_2}
\footnotesize  
\renewcommand{\arraystretch}{1.2}
\rowcolors{2}{lightgray!30}{white}
\begin{tabular}{%
    >{\raggedright\arraybackslash}p{3.7cm}
    >{\centering\arraybackslash}p{0.9cm}
    >{\centering\arraybackslash}p{1.2cm}
    >{\centering\arraybackslash}p{1cm}
    >{\centering\arraybackslash}p{2.1cm}
    >{\centering\arraybackslash}p{1.0cm}
    >{\centering\arraybackslash}p{1.4cm}
    >{\centering\arraybackslash}p{1.4cm}
    >{\centering\arraybackslash}p{1.4cm}}
\toprule
\rowcolor{gray!30} \textbf{Disease Category} & \multicolumn{3}{c}{\textbf{PoPStat}} & \textbf{CI}& \textbf{PoPStat} & \multicolumn{3}{c}{\textbf{Closest countries to the reference}} \\
\rowcolor{gray!30} & {r} & {Reference country} & {p} &  &\textbf{ r$^2$} & {First} & {Second} & {Third} \\
\midrule 
\multicolumn{9}{l}{\textbf{Level 3 Disease Categories}} \\
Ischemic heart disease & -0.75 & Latvia & < 0.001 & (-0.807, -0.678) & 0.562 & Estonia & Lithuania & Slovenia\\
Stroke & -0.497 & France & < 0.001 & (-0.599,  -0.379) & 0.247 & Belgium & United Kingdom & Finland\\
Pulmonary Arterial Hypertension & 0.375 & Oman & < 0.001 & (0.242,  0.494) & 0.14 & Saudi Arabia & Maldives & Bahrain\\
Chronic obstructive pulmonary disease & -0.621 & France & < 0.001 & (-0.704, -0.523) & 0.386 & Belgium & United Kingdom & Finland\\
Asthma & -0.594 & Central African Republic & < 0.001 & (-0.681, -0.49) & 0.352 & Niger & Somalia & Chad\\
Breast cancer & -0.815 & Japan & < 0.001 & (-0.859,  -0.759) & 0.664 & Italy & Portugal & Greece\\
Colon and rectum cancer & -0.922 & Japan & < 0.001 & (-0.941,  -0.897) & 0.85 & Italy & Portugal & Greece\\
Cervical cancer & -0.257 & Tonga & < 0.001 & (-0.388, -0.116) & 0.066 & Samoa & Syria & Comoros\\
Prostate cancer & -0.716 & Japan & < 0.001 & (-0.781,  -0.637) & 0.513 & Italy & Portugal & Greece\\
Liver cancer & 0.499 & Oman & < 0.001 & (0.381,  0.601) & 0.249 & Saudi Arabia & Maldives & Bahrain\\
Cirrhosis and other chronic liver diseases & 0.162 & Qatar & 0.03 & (0.016,  0.3) & 0.026 & United Arab Emirates & Bahrain & Maldives\\
Inflammatory bowel disease & 0.634 & Oman & < 0.001 & (0.537,  0.714) & 0.401 & Saudi Arabia & Maldives & Bahrain\\
Alzheimer's disease and other dementias & -0.925 & Japan & < 0.001 & (-0.943,  -0.9) & 0.855 & Italy & Portugal & Greece\\
Parkinson's disease & -0.936 & Japan & < 0.001 & (-0.952, -0.916) & 0.877 & Italy & Portugal & Greece\\
Alcohol use disorders & 0.438 & Oman & < 0.001 & (0.312,  0.549) & 0.192 & Saudi Arabia & Maldives & Bahrain\\
Diabetes mellitus & -0.355 & Sri Lanka & < 0.001 & (-0.476,  -0.22) & 0.126 & Saint Vincent and the Grenadines & Brazil & Argentina\\
Chronic kidney disease & -0.411 & Uruguay & < 0.001 & (-0.525,  -0.282) & 0.169 & New Zealand & Iceland & Australia\\
Rheumatoid arthritis & -0.713 & Puerto Rico & < 0.001 & (-0.778,  -0.633) & 0.509 & Portugal & Croatia & Finland\\
Maternal disorders & -0.909 & Central African Republic & < 0.001 & (-0.932,  -0.88) & 0.827 & Niger & Somalia & Chad\\
Neonatal disorders & -0.91 & Central African Republic & < 0.001 & (-0.932, -0.881) & 0.827 & Niger & Somalia & Chad\\
Self-harm & 0.443 & United Arab Emirates & < 0.001 & (0.318,  0.553) & 0.197 & Qatar & Bahrain & Maldives\\
Interpersonal violence & -0.472 & Tonga & < 0.001 & (-0.578,  -0.351) & 0.223 & Samoa & Syria & Comoros\\
HIV/AIDS & -0.549 & Central African Republic & < 0.001 & (-0.644,  -0.439) & 0.302 & Niger & Somalia & chad\\
Tuberculosis & 0.759 & Malta & < 0.001 & (0.689, 0.815) & 0.576 & Poland & Cyprus & Slovakia\\
Dengue & -0.578 & Jamaica & < 0.001 & (-0.687,  -0.444) & 0.334 & Colombia & Saint Lucia & Brunei\\
Protein-energy malnutrition & 0.561 & Malta & < 0.001 & (0.452,  0.654) & 0.315 & Poland & Cyprus & Slovakia\\
\bottomrule
\end{tabular}
\end{table}

\section{Discussion}
\subsection{PoPDivergence and PoPStat}
In this study, we devised two novel metrics, PoPDivergence and PoPStat. PoPDivergence measures the difference between the shape of a country’s population pyramid and that of a reference population pyramid. PoPStat is the correlation between PoPDivergence and the (natural log form of) death rate for a disease of interest.  We also designed an optimization mechanism to select a reference population that invariably maximizes this correlation. PoPStat, in other words, is a measure that tells you how tightly a specific disease is associated with the population pyramid. To the best of our knowledge, this is the first study to have introduced a metric that quantifies this association. PoPStat also provides information on the nature of population structures most or least compatible with high mortality from that specific disease.

Meaningfully summarizing the multi-dimensional population structure into a scalar variable allows us to explore its relationship with various outcomes. Though the current study has focused on disease mortality, the correlation with PoPDivergence can be assessed for any social, behavioural or economic phenomena of choice. On the exposure end, a range of causal information is captured by the shape change in the population pyramid. Firstly, the shift in sex and age composition between pyramids directly affects gender- and age-specific mortality rates of a specific cause. Secondly, despite being a direct function of the rate of birth, deaths and migration, the population pyramid also embodies multiple biological, socio-economic, behavioral and cultural attributes that directly or indirectly compound disease-specific mortality. Thirdly, the shape of population pyramids also carries a wider representation of the geographic or regional characteristics that may impact global mortality patterns.

One major challenge to studying the dynamics of population change has been the resource-intensiveness of analyzing data not only across geographies but also across generations. But, each country and region is at a unique point in its demographic transition as reflected by their population pyramids. Therefore, PoPDivergence allows us to capture the longitudinal axis of global population change within a single cross-section in time. The observation that the reference population mostly happens to be at either end of the demographic transition pyramid is also noteworthy. For any given disease category, with increasing PoPDivergence, we see a transition of population pyramids similar to that described by the demographic transition model without any explicit instructions to follow this path. So, any monotonic metric linked to PoPDivergence will allow us to empirically visualize the epidemiological, social, and economic transitions that happen parallel to the demographic transition.

\subsection{Performance of PoPStat with other indicators}
Examining IHD as an illustrative example, the association of IHD mortality with PoPDivergence - represented by the PoPStat - was superior to the correlations of other demographic indicators included in this study. This pattern remained true for all the relevant causes and cause clusters across the three levels (Supplemental File 1). This indicates that other scalar variables such as median age and life expectancy may condense lesser information about the demography and its health compared to the PoPDivergence. Therefore, PopDivergance becomes a high-quality dimension reduction of the information-rich population pyramid. For example, in relation to Figure \ref{fig3:median_index}, PoPDivergence captures the shape differences of the two population pyramids that share a singular median age. However, it must be noted that the correlation between median age and disease mortality remained equally strong for a majority of causes (Supplemental File 1). So, future research can explore if the ability of these two indicators to associate with the underlying demographic structures of the demographic transition remains true for areas other than epidemiology.

PoPStat also performed better than the disease mortality correlations of other development-related indicators. It was noteworthy that PoPStat had a better association with the social determinants of IHD mortality than HDI which was a latent variable composite of life expectancy, years of schooling and Gross National Income (GNI) per capita. It was also superior to the correlation between IHD mortality and GDI per capita taken alone. However, it has to be noted that the relationship between health and the national income of a country is likely curvilinear - similar to Preston’s curve\cite{T_H._1975}. Pearson’s correlation may not do justice in analyzing such an association. Nevertheless, these findings demonstrate the versatility of PoPDivergence as a measure of the population structure during disease burden assessments. It does not only associate more strongly with demographic determinants of mortality compared to other population descriptors but also aligns more closely with socio-economic determinants of mortality than conventional development metrics. 

\subsection{Level 1 Disease categories}
The higher PoPStat of NCDs indicated that they are more strongly associated with the underlying population pyramid of a country and its latent constructs than CMNN diseases. This disparity may have arisen due to intrinsic disease attributes, such as chronic diseases being selective for older ages while infections cross-cut age strata. It may have also been due to the pooling of heterogeneous entities like communicable, maternal, neonatal and nutritional causes, which may have diluted the measure of association. The lowest PoPStat was noted for injuries. The largest contribution to injuries is made from unintentional injuries such as falls, drowning, fire and natural disasters whose risk may similarly spread across several age cohorts\cite{T_Chandran_2010}. This may have resulted in the poor association of PoPDivergence with external causes of death compared to natural causes of death that originate from disease processes of ageing. Under-reporting of the true burden of injuries by national surveillance systems may have also contributed to the poor correlation with the countries’ demographics.

The \textbf{Epidemiological Transition Model} introduced by Abdel Omar in 1971 expanded Landry’s DTM by incorporating disease and mortality patterns\cite{T_R._2005}. One major proposition in ETM was that parallel to the demographic transition, infectious diseases would be gradually replaced by degenerative and man-made diseases as the leading cause of death. However, the validity of ETM has been heavily contested in literature due to the linearity in its transition and the many deviations observed. Santhosa et al. (2014) in their review of literature over four decades identified a discordance between the theory of ETM and empirical evidence\cite{T_Santosa_2014}. Therefore, the observation of NCD mortality being associated with constrictive population pyramids whereas the burden of deaths due to CMNN diseases being linked to expansive pyramids critically adds to the current discourse on ETM. Our findings indicate that when the demographic stages of all the countries are captured within a single timeframe, the propositions of ETM may hold water even after half a century. However, it must be noted that PoPDivergence cannot assess the pace of these transitions and cannot identify which ET model: classical, accelerated or delayed, fits the contemporary global picture best.

Considering raised NCD mortality among regressive population structures, in 2019, 18777 Years of Life were Lost (YLL) due to NCDs in very high Human Development Index (HDI) countries compared to 7444 YLL in very low HDI countries\cite{G_Emadi_2021}. Omar identified socio-economic, political, cultural, ecological and health system determinants for this epidemiological transition\cite{T_R._2005}.  First, the pathological processes underlying the most fatal NCDs - cardiovascular diseases and cancers - are closely associated with the age structure, making them highly prevalent among aging populations\cite{B_Lv_2024}. Next, the crude prevalence and attributable burden of primary risk factors for NCDs such as physical inactivity and alcohol consumption; and intermediate risk factors like hypertension were also greater in the GBD regions:  High Income North America, High Income Asia Pacific, Western, Central and Eastern Europe, with predominantly constrictive population pyramids\cite{P_T_2022,A_G_2018,G_H._2017}. Even after adjusting for age, the prevalence and attributable burden of risk factors like tobacco smoking, alcohol use, dietary risks and obesity remained elevated for most of these GBD regions\cite{S_B_2021,T_Degenhardt_2018,T_Chong_2023}. This indicates that the strong PoPStat for NCDs may be due to capturing latent information within population pyramids that extend beyond age and sex dynamics.

Considering environmental exposures associated with the urbanization and industrialization complex, these regions with constrictive pyramids interestingly had a lower attributable burden of disease due to air pollution. However, the attributable cancer mortality due to occupational carcinogens was contrastingly higher than in other populations\cite{E_J_2017,G_undefined_2020}. These are the likely risk exposures that have led to a higher NCD burden in regressive population pyramids. Finally, health system factors also explain the burden of NCDs in constrictive population pyramids. Advanced health technologies and higher per capita health expenditures may have contributed to longevit	y and the consequent chronic health problems of those aging populations.

The relationship between communicable diseases and expansive population pyramids also operates on multiple levels. First, communicable diseases with the highest mortality - for example, enteric infections - mainly affect younger populations. Therefore, the age structure of broad-based population pyramids has a direct impact on infectious disease mortality\cite{E_Troeger_2018}.  Next, the burden of malaria, tuberculosis, HIV/AIDs and neglected tropical diseases is primarily held by countries with expansive population pyramids in Africa, Asia and Latin America\cite{G_J_2024}. This may be due to multiple socio-economic determinants such as poverty and colonial past that are shared by them or due to common ecological characteristics such as temperate climate that facilitate the survival of specific pathogens and vectors\cite{P_Zhu_2024}.

The association between injuries and expansive population pyramids has been previously documented in the literature. For example, in 2021, age-standardized Disability Adjusted Life Years (DALY) rates for injuries among males were greater in the low Social Development Index (SDI) countries (5958.8 per 100,000) compared to the high SDI countries (3050.8 per 100,000)\cite{G_J_2024}. This may be attributed to low-quality infrastructure, increased occupational hazards, lack of safety standards, limited injury prevention and disaster preparedness efforts in countries at the early stages of demographic transition\cite{T_Chandran_2010}.

\subsection{Level 2 Disease Categories}
\subsubsection{Non Communicable Diseases}
Considering Level 2 NCD categories, the highest PoPStat was seen among neurological disorders. This is understandable as the largest mortality due to neurological disorders is caused by strokes and degenerative conditions like Alzheimer’s disease which are mostly associated with older age and thereby the demographic structure\cite{G_L_2019}. The PoPStat of neurological disorders was even greater than cancers which tend to manifest in children and young ages despite their high overall burden.  However, following age standardization, regions with expansive population pyramids also showed an increased number of DALYs indicating that the association of neurological disorders to population pyramids was largely mediated by age\cite{G_L_2019}.

Association of mental disorders and constrictive population pyramids aligned with the existing knowledge base as the highest number of DALYs due to mental diseases were observed among populations in North America, Australia and Western Europe that were in the latter stages of the demographic transition\cite{G__2022}. Substance use disorders being linked to constrictive population pyramids was also predictable given the most used substance was alcohol which had the largest crude prevalence in Western and East European regions and North America after the Asian region\cite{T_Degenhardt_2018}. Since alcohol and drug use was not limited to an age stratum\cite{T_Degenhardt_2018}, in relation to substance use disorders, PoPStat did not only correlate with the age structure but also reflected socio-cultural determinants that these populations may have acquired during their demographic transition.

Chronic obstructive pulmonary disease (COPD) is the main cause of chronic respiratory disease mortality worldwide\cite{G_Momtazmanesh_2023}. Chronic respiratory diseases being connected to less-inverted population pyramids is explainable on this basis where a higher burden of COPD deaths was observed among Western European population pyramids with comparable shapes\cite{G_Momtazmanesh_2023}. However, when adjusted for age, the highest death rates of chronic respiratory diseases were seen among African and Asian regions with expansive population pyramids. This indicates that  - similar to neurological disorders - the association of chronic respiratory diseases with the population pyramid was mediated by the age structure. The reduced burden of digestive disease mortality among asymmetrically constrive population pyramids in the Middle East was noteworthy.  Given the second largest cause of digestive disease deaths is cirrhosis, this may be attributed to the cultural characteristics of such population pyramids where alcohol consumption is limited\cite{G_Wang_2023,T_Degenhardt_2018}.

Diabetes was the sole exception to the relationship between NCD mortality and regressive population pyramids observed in this study. Higher diabetic mortality among expansive population pyramids was in line with the current evidence where the death rates were highest in the GBD super regions of Latin America and the Caribbean, North Africa and the Middle East, Sub-Saharan Africa and South Asia\cite{G_Liane_2023}. This deviation could be due to a myriad of causes:  high BMI or obesity in regions such as North Africa and the Middle East,  genetic predispositions among regions such as South Asia and determinants such as poverty, limited access to healthy food and lack of comprehensive diabetic care that are spread across these regions\cite{G_Liane_2023}. It is also important to note that diabetes carried the lowest PoPStat among NCDs. This suggests that the contribution of some of the above factors may not be captured by the population pyramid. However, compared to other demographic and developmental indicators, PoPStat still had the best correlation with diabetic mortality (Supplemental File 1).

\subsubsection{Communicable Diseases}
NTDs had the highest PoPStat among the Level 2 communicable diseases and the majority of the burden was shared by expansive population pyramids. NTDs are a heterogeneous group of parasitic, bacterial and viral infectious diseases that commonly occur in tropical regions, predominantly among low and middle-income countries of Asia, Africa and Latin America. However, between 1990 and 2019 the most pronounced increase in NTD incidence was observed among the middle and high SDI regions\cite{G_Lin_2022}. Our findings indicate that the spread of NTDs to developed populations away from the equator has not diluted their association with progressive population pyramids in the tropics. This may be due to haphazard urbanization and poor waste management of growing populations that support vector breeding, limited public health infrastructure including sanitation and other geographical and environmental attributes shared by these populations\cite{G_Lin_2022}. However, the correlations of NTD mortality with UHC and HDI were seen to be higher than PoPStat (Supplemental File 1). This suggests that the impact of factors such as primary health care delivery and education may be larger on NTDs than direct and indirect demographic influence.

Other infectious diseases had a strong PoPStat, with the major mortality seen among expansive population pyramids. This was partly due to the largest burden of other infections being caused by meningitis - a condition frequently encountered among expansive population pyramids in the African region\cite{G_L_2019}. However, the correlation of the mortality of other infectious diseases with UHC was also higher than PoPStat implying that health system factors may play a greater role in communicable disease mortality than demographic variables. HIV and other STIs had a moderate PoPStat and the burden of deaths reduced in the shift from progressive to regressive demographic structures. HIV could be the main culprit behind this finding as it made the largest contribution to this Level 2 category with the highest mortality rates seen among expansive populations in the Sub-Saharan African region\cite{G_Carter_2024}. Respiratory infections had the lowest PoPStat among communicable diseases. This was likely due to respiratory infections and their causative pathogens affecting both sexes in all age strata and thereby not being discriminated by the sex-age structure. However, further investigations are needed to see why respiratory infections were low among asymmetrically constrictive pyramids such as Middle Eastern populations. This may have been due to the dry climate, lack of seasonal changes and other regional characteristics shared by them than due to an intrinsic feature of the unique shape.

\subsubsection{Maternal, neonatal and nutritional diseases}
Among other CMNN categories, maternal and neonatal conditions showed a very strong PoPStat. This tight association with the demography may seem self-evident as a higher number of mothers and neonates in broad-based populations may result in a higher burden of maternal and neonatal deaths. Nevertheless, even after adjusting for age, the greatest DALY rates due to most neonatal causes were from low SDI countries in Asia and Africa and the highest DALY rates due to all maternal causes were from Sub-Saharan Africa. Meanwhile, the lowest age-standardized DALY rates due to all neonatal and maternal causes were from high-income Asia Pacific, North America and Australasia regions\cite{G_Peng_2024}. So, the reduction in mortality due to maternal and neonatal causes when transitioning from progressive to regressive population structures is not merely due to changes in the pyramid base but is also influenced by socio-economic and health determinants acquired during the process. Nutritional deficiencies carrying a moderate PoPStat was interesting given death rates due to malnutrition being strongly polarized between low SDI (11.5 per 100000) and high SDI (0.6 per 100000) regions in a manner similar to maternal and neonatal causes\cite{T_Chong_2023}. WHO’s Eastern Mediterranean region showing a mixed picture with higher death rates due to both malnutrition and obesity could have dampened the PoPStat of nutritional-related mortality\cite{T_Chong_2023}. 

\subsubsection{Injuries}
Higher mortality of transport injuries among expansive population pyramids tallied with the literature. Globally, the highest DALYs due to TIs were reported between 20 to 40 years old indicating a direct age-related phenomenon\cite{G_J_2024}. After age standardization, the bulk of DALY rates due to TIs were still noted among low SDI regions with rising trends in Western Sub-Saharan African, Southern Latin American, South and East Asian populations \cite{G_Wan_2023}. This meant that in addition to the accident-prone age groups, environmental determinants such as unsafe roads, cultural determinants like the use of motorbikes and tuk-tuks and political determinants such as limited enforcement of traffic laws may be layered within the associated population pyramids.  Interestingly, it has been recognized that TI incidence had a positive correlation with SDI in most of the above regions\cite{G_Wan_2023}. This suggests that certain dimensions of the development transition such as urbanization and motorization may themselves increase TI mortality. This could have also attenuated the original PoPStat for transport injuries.

An interesting PoPDivergence emerged when examining the mortality of unintentional injuries. The reference population of Kuwait was relatively in the middle of the demographic transition process while having a lower burden of UI mortality. As KL divergence happens away from this reference it would lead to either predominantly young or predominantly old populations. Therefore, based on the optimization followed in this study, higher UI mortality appears to be associated with populations at either extremities in age. This made sense given the largest burden of UI was due to falls and drowning where both children and older persons carried the highest risk\cite{G_J_2024}.

Self-harm and interpersonal violence demonstrated a very weak PoPStat. This was noteworthy given the increase in self-harm during the development transition has been theorized early in sociology. Emilie Durkheim who studied suicides in the latter part of the 19th century proposed that self-harm was governed by the changes of social integration during the shift of societies from mechanistic to organic solidarity\cite{S_Durkheim_2005}. However, our findings indicate that self-harm and violence may not make such a strong discrimination between developing and developed populations. Though both conditions shared a higher mortality burden among expansive population pyramids than constrictive ones, 98\% of the variance in mortality could not be explained by the said transition. This is also empirically supported by GBD studies where high-SDI countries had the second highest age-adjusted mortality rate (11.1 per 100000) after low-middle SDI countries (11.9 per 100000) attributed to self-harm\cite{G_Zhou_2024}. Though risk factors such as intimate partner violence, high temperatures and socio-political unrest may explain interpersonal violence among expansive population pyramids in the tropical regions, self-harm seems to affect all populations irrespective of their position in the demographic change\cite{G_Zhou_2024}.

\bibliographystyle{vancouver}
\bibliography{references.bib}

\end{document}